\documentclass[11pt]{iopart}
\usepackage{iopams}
\usepackage{graphicx}
\usepackage{braket}
\usepackage{xcolor}
\usepackage[utf8x]{inputenc}
\usepackage{float}
\usepackage{bm}
\usepackage[T1]{fontenc} 

\begin{document} 

\title{Asymmetric Light Scattering from an Atomic System with Gain: A Quantum Analysis}

\author{Lorena Acevedo}
\address{Departamento de F\'isica Te\'orica, At\'omica y \'Optica,  Universidad de Valladolid, 47011 Valladolid, Spain}
\author{Manuel Donaire}
\address{Departamento de F\'isica Te\'orica, At\'omica y \'Optica,  Universidad de Valladolid, 47011 Valladolid, Spain}
\ead{manuel.donaire@uva.es}
\begin{abstract}
  We study the scattering of light by a binary system of identical atoms in which one of them is incoherently pumped. This system belongs to the kind of non-parity symmetric optical systems in which gains and losses are partially compensated. 
We carry out a fully quantum analysis of  the directionality of the radiation scattered from the atoms when the incident light strikes the system either perpendicular or alongside the interatomic axis. By resolving the emitted field in space, the microscopic mechanisms associated with photon exchange, interference, and reciprocity become directly identifiable. We find that, generally, while the degree of asymmetry  depends on the pump rate, the preferred direction for emission depends on the interatomic distance and the detuning of the probe field with respect to the resonant frequency. On physical grounds,  for the case of frontal illumination, the asymmetry is the result of the interference of the photons emitted from different atoms. On the contrary, for side lighting,  the asymmetry with respect to the side of incidence is caused by both the phase difference between the probe field photons that strike each atom and the interference between the photons emitted from different atoms. Further, for side lighting too, our quantum approach demonstrates that the forward scattered power depends on the side of incidence, which reveals the lack of reciprocity in the quantum optical response of the system. This result conflicts with what is obtained within a classical approach. 
\end{abstract}
\begin{small}
\end{small}
\maketitle

\section{Introduction}\label{sec:intro}

Understanding how light is redistributed after interacting with a quantum emitter is essential for controlling photon transport in quantum optical platforms. Beyond integrated quantities such as extinction or absorption, the angular distribution of the scattered radiation contains detailed information about interference, phase accumulation, and the pathways followed during the scattering process. These directional properties are particularly relevant for engineered atomic systems in which coherent interactions compete with gain or dissipation, opening possibilities for tailoring emission patterns and realizing nonreciprocal optical functionalities. 

Recently, considerable attention has been devoted to optical systems combining active and passive elements. For instance, effectively non-Hermitian optical systems are made of an arrangement of active and passive optical elements. In classical photonic structures, the interplay between gain, loss,
and interference has enabled asymmetric scattering, unidirectional transport, and
scattering suppression \cite{Vincenzo,scatotros}, which have been exploited in applications \cite{la36MJ,la34MJ,la38MJ,sensors,antenas,photovolts}. This is for instance the case of classical metallic dimers with balanced gains and losses \cite{ManjavacasPT1,ChenetalPRA,ManjavacasSandersPT2}, in which the strong coupling of surface plasmons and light is utilized.  In particular, Ref.\cite{ManjavacasPT1} analyses the anisotropic features of the optical response of these systems  implementing the gains in an effective manner by means of a Lorentzian term in the dielectric function of the active elements.


Similar ideas have motivated the study of atomic implementations, where the microscopic dynamics can be described entirely within quantum electrodynamics and individual scattering processes can be resolved explicitly. Hence, 
it is the purpose of this article to analyse the directionality of scattering by a system made of two identical atoms in which one of them is excited by an incoherent pump.  In a previous publication, Ref.\cite{Acevedo}, the analysis involved the global optical properties of the system. 
There, it was shown that losses and gains are partially compensated, which leads to a reduction of the total extinction cross-section, but not quite to its full vanishing characteristic of  $\mathcal{PT}$-symmetric systems \cite{ManjavacasPT1,ManjavacasSandersPT2,la36MJ,Vincenzo,Japs,ORN19,JKP16}. Although that quantity revealed how pumping modifies the global optical response, it does not determine how the scattered radiation is redistributed in space. The purpose of the present work is to address this complementary question. Rather than investigating integrated cross sections, we analyze the angular distribution of the scattered field and identify the microscopic origin of directional emission, elucidating to what extent the features of the  asymmetric response found in analogous classical systems are indeed generic \cite{ManjavacasPT1}. Particular attention is devoted to the role of interference between different scattering pathways, the influence of interatomic separation, and the dependence on the side from which the probe field illuminates the system. These considerations naturally lead to the analysis of parity asymmetry and optical reciprocity within the same microscopic framework. Our approach retains the advantages of the wave-function perturbative formalism, allowing each contribution to the scattered radiation to be associated with an individual quantum process. This microscopic description provides direct physical insight into the origin of asymmetric scattering while preserving the full quantum character of the atom-field interaction.



We find that the features of the asymmetric response are not generic since, generally, the scattering imbalance towards one atom or the other depends on the interatomic distance as well as on the detuning of the probe field with respect to the resonant frequency. Further, while the classical optical response of the atomic system is reciprocal, which manifests in the independence of the forward-scattered power with respect to the side of incidence, its quantum optical response is non-reciprocal. This is interpreted on the basis of the violation of parity and time-reversal symmetries.


The article is organized as follows. In Sec.~\ref{lasec2} we outline the fundamentals of the formalism. Sec.~\ref{lasec3} contains the classification of the scattering processes together with general considerations about the calculation of emission powers. In Sec.~\ref{lasec4} we analyse the case of frontal illumination, while in Sec.~\ref{lasec5} we deal with side lighting. In Sec.~\ref{lasec6} we carry out an analogous classical calculation. We finalize with the discussion of the results in Sec.~\ref{lasec7}, together with a comparison of our results with those obtained from the classical approach.
	
\section{Methods and fundamentals}\label{lasec2}	

In this Section we outline the fundamentals of our approach. We consider two identical  atoms separated by a distance $R$, with one atom incoherently
pumped and both atoms illuminated by a weak probe field. The complete derivation of the steady-state dynamics
and the implementation of the incoherent pump are described in detail in Refs.\cite{Acevedo,myPRA}. However, the present work addresses
a different observable: the angular distribution of the scattered radiation and its dependence on the
direction of incidence. We therefore retain only the elements of the previous formalism required to
calculate the differential scattered power.

\subsection{Mixed atomic state}
\begin{figure}
	\begin{center}
		\includegraphics[width=110mm,angle=0,clip]{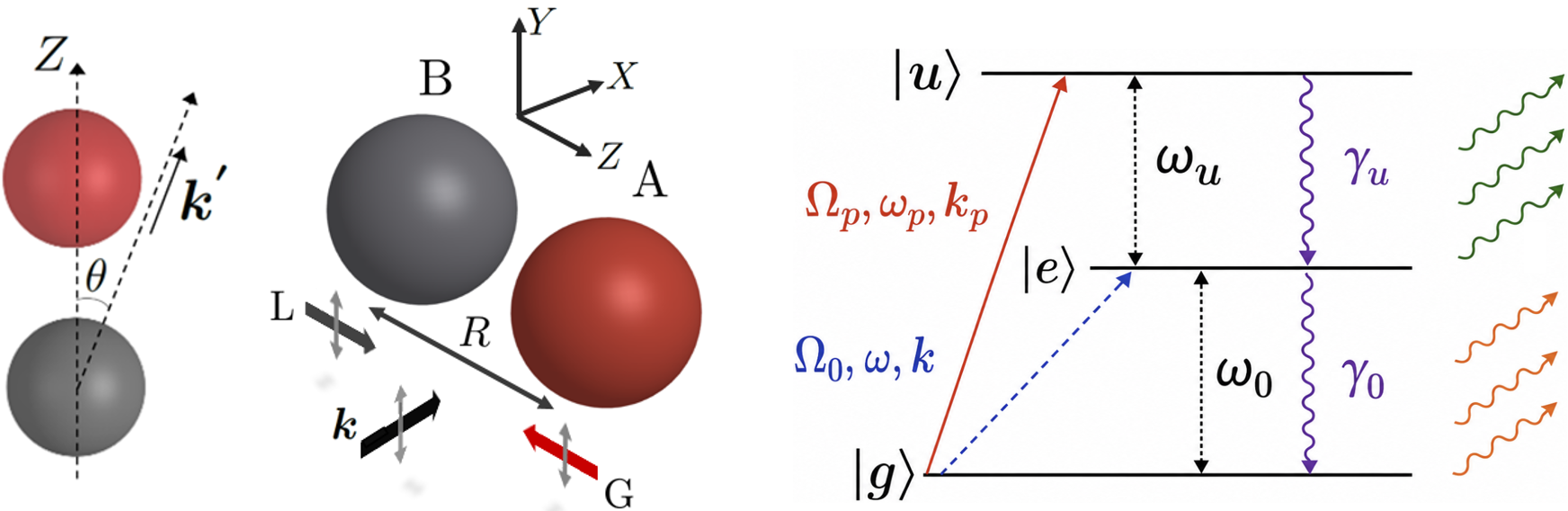}
		\caption{Left panel: Schematics of the two-atom system illuminated by a probe field of  polarization transverse to the interatomic axis and momentum \textbf{k}, quasi-resonant 
			with the $g\rightleftarrows e$ transition. Incidence is either transverse or alongside the interatomic axis. Radiation of momentum $\bf{k'}$ is scattered at an angle $\theta$ with respect to the interatomic axis. Right panel: Energy spectrum of the atoms. Levels $g$ and $e$ are the ground and excited states of the effective two-level atoms. In the active atom, $A$, a pump field of strength  $\Omega_{p}$ causes the transient excitation from $g$ to the auxiliary state $u$, which decays spontaneously into $e$ at a rate $\gamma_{u}$. This originates the effective and  incoherent pump from state $g$ to $e$ at a rate $\mathcal{P}=\Omega_{p}^{2}/\gamma_{u}$ \cite{Acevedo,myPRA}.} \label{fig1}
	\end{center}
\end{figure}

The two atoms of our system, say $A$ and $B$,  behave as two-level ones, with ground state $g$ and excited state $e$, $\omega_{0}$ being the transition frequency between them and $\gamma_{0}$ being the decay rate in free-space. While one of the atoms, say $B$, remains in its ground state, atom $A$ is continuously and incoherently pumped through an auxiliary state --cf. Ref.\cite{myPRA} and energy scheme of Fig.\ref{fig1}. Hereafter atoms $A$ and $B$ will be referred to as active and passive atom, respectively. Besides, both atoms are continuously illuminated with a quasi-resonant probe field of frequency $\omega\approx\omega_{0}$, Rabi frequency $\Omega_{0}\ll\gamma_{0}$, and density of photons $N_{\mathbf{k},\boldsymbol{\epsilon}}$ with momentum $\mathbf{k}$ and polarization vector $\boldsymbol{\epsilon}$. As a result, the steady state of the system is made of the tensor product of the ground state of atom $B$, the $N_{\mathbf{k},\boldsymbol{\epsilon}}$-photon state of the probe field, and a statistical  mixture of the pure states $\tilde{g}$ and $\tilde{e}$ of the active atom --the 'tilde' here denotes that the states belong to a statistical mixture, i.e., $|\Psi_{0}\rangle=|\Psi_{0}\rangle_{g}+|\Psi_{0}\rangle_{e}$ with
\begin{equation}
	|\Psi_{0}\rangle_{g}=\sqrt{\gamma/\Gamma}\:|N_{\mathbf{k},\boldsymbol{\epsilon}}\rangle\otimes|\tilde{g}\rangle\otimes|g\rangle,\quad
	|\Psi_{0}\rangle_{e}=\sqrt{\mathcal{P}/\Gamma}\:|N_{\mathbf{k},\boldsymbol{\epsilon}}\rangle\otimes|\tilde{e}\rangle\otimes|g\rangle.
	\label{wavefunction}
\end{equation}
In these equations $\gamma$ is the total decay rate of the transition $e\rightarrow g$, which contains the sum of the natural decay $\gamma_{0}$ and the non-radiative decay rate $\gamma_{nr}$, $\mathcal{P}$ is the pump rate, and $\Gamma=\gamma+\mathcal{P}$. 
Throughout this work this mixed state serves as the initial condition for the perturbative calculation of the scattering amplitudes.

For the sake of simplicity we will assume that the dipole transition moment,  $\boldsymbol{\mu}=\langle g|\mathbf{d}|e\rangle$, with $\mathbf{d}$ being the electric dipole operator,  lies on the plane perpendicular to the interatomic axis. In turn, this implies an equal contribution of near field and far field interatomic interactions to the optical response of the system. Considering the interatomic axis along the cartesian $z$-axis, we take $\boldsymbol{\mu}=\mu(\hat{\mathbf{x}}+\hat{\mathbf{y}})/\sqrt{2}$. Further, we will distinguish two cases for the direction of incidence of the probe field, namely, perpendicular and parallel to the interatomic axis, with wave vectors $\hat{\mathbf{k}}=\hat{\mathbf{x}}$ and $\hat{\mathbf{k}}=\pm\hat{\mathbf{z}}$, respectively, and polarization vector $\boldsymbol{\epsilon}=\hat{\mathbf{y}}$ in both cases --see left panel of Fig.\ref{fig1}.

\subsection{Atoms-field interaction. Coherent evolution of the system}

Our system is broadly speaking an open system due to the aforementioned integration of the pump field which, in turn, leads to the incoherent evolution of the two-atom system that we describe in the next subsection. However, in regards to the evolution the atoms induced by the probe field, this can be treated unitarily within the framework of time-dependent perturbation theory and the wavefunction formalism. The atoms-EM field system evolves according to the time-propagator $\mathbb{U}(t)$
\cite{Sakurai},
\begin{equation}
	\mathbb{U}(t-t_{0})=\mathcal{T}\textrm{exp}\left\{-i\hbar^{-1}\int_{t_{0}}^{t}d\tau\:H(\tau)\right\},
\end{equation}
where $H$ is the total Hamiltonian of the system which is made of a free component, $H_{0}$, 
\begin{equation}
	H_{0}=\sum_{i=A,B}\hbar\omega_{0}\ket{e}_{i}\bra{e}_{i}+
	\sum_{\mathbf{k}',\boldsymbol{\epsilon}'}\hbar\omega'(a^{\dagger}_{\mathbf{k}',\boldsymbol{\epsilon}'}a_{\mathbf{k}',\boldsymbol{\epsilon}'}+\frac{1}{2}),\nonumber 
\end{equation}  
and an interaction Hamiltonian, $W$, that accounts for the atom-field interaction in the electric dipole approximation and is a perturbation with respect to $H_{0}$,
\begin{equation}
	W=-\sum_{i=A,B}\mathbf{d}_{i}\cdot\mathbf{E}(\mathbf{r}_{i}).\nonumber
\end{equation}
In theses equations, $a_{\mathbf{k}',\boldsymbol{\epsilon}'}^{\dagger}$ and 
$a_{\mathbf{k}',\boldsymbol{\epsilon}'}$ are the creation and annihilation operators of photons of momentum  $\mathbf{k}'$, frequency $\omega'=c\:k'$ and polarization vector $\boldsymbol{\epsilon}'$, respectively; $\mathbf{r}_{i}$ is the centre of mass of each atom, $i=A,B$; and the electric field operator can be written as a sum over normal modes,
\begin{equation}
	\fl\mathbf{E}(\mathbf{r})=i\sum_{\mathbf{k}',\boldsymbol{\epsilon}'}\sqrt{\frac{\hbar\omega'}{2\epsilon_{0}\mathcal{V}}}
	\left[\boldsymbol{\epsilon}'a_{\mathbf{k}',\boldsymbol{\epsilon}'}e^{i\mathbf{k}'\cdot\mathbf{r}}-\boldsymbol{\epsilon}'^{\ast}a^{\dagger}_{\mathbf{k}',\boldsymbol{\epsilon}'}
	e^{-i\mathbf{k}'\cdot\mathbf{r}}\right]=\sum_{\mathbf{k}',\boldsymbol{\epsilon}'}\left[\mathbf{E}^{(+)}_{\mathbf{k}',\boldsymbol{\epsilon}'}(\mathbf{r})+\mathbf{E}^{(-)}_{\mathbf{k}',\boldsymbol{\epsilon}'}(\mathbf{r})\right].\label{fieldE}
\end{equation}

The perturbative approach involves the expansion of the propagator $\mathbb{U}(t-t_{0})$  in powers of  $W$ from its
time-ordered exponential expression,
\begin{equation}
	\mathbb{U}(t-t_{0})=\mathbb{U}_{0}(t)\:\mathcal{T}\textrm{exp}\int_{t_{0}}^{t} -i\hbar^{-1}\mathop{d\tau} \mathbb{U}_{0}^{\dagger}(\tau)W
	\mathbb{U}_{0}(\tau-t_{0}),\label{propagator}
\end{equation}
where $\mathbb{U}_{0}(t-t')$ is the unperturbed time-propagator, $\mathbb{U}_{0}(t-t')=\exp{[-i\:\hbar^{-1}H_{0}(t-t')]}$. In particular, we will be interested in scattering processes in which the statistical mixture of states $|\Psi_{0}\rangle_{g,e}$ is propagated in time  under the action $\mathbb{U}(t-t_{0})$ from the initial time $t_{0}=0$  up to the observation time $t$, at which the final state contains a scattered photon --see the diagrams of Fig.\ref{fig2}. Since we will be interested in the angular distribution of the scattered radiation, only a sum over polarization states will be carried out over the normal modes of that scattered photons,
\begin{equation}
	\sum_{\boldsymbol{\epsilon}_{k}}\boldsymbol{\epsilon}_{k}\otimes\boldsymbol{\epsilon}_{k}^{\ast}=\frac{\mathbf{k}\otimes\mathbf{k}}{k^{2}}.
\end{equation}
In contrast, the intermediate states of each scattering processes will contain virtual photons which play two complementary roles. On the one hand, they mediate the dipole-dipole interaction between both atoms. On the other hand, they mediate the induction of a dipole moments on each atom by the electric field sourced by the other one.  The virtual photons of those intermediate states contribute with any polarization and momentum direction, reason why a sum over polarization states and orientations is to be performed. In the perturbative calculations this involves the evaluation of the vacuum expectation value of the quadratic fluctuations of the electric field,
\begin{equation}
	\sum_{\mathbf{k}',\boldsymbol{\epsilon}'}\bra{0}\mathbf{E}^{(+)}_{\mathbf{k}',\boldsymbol{\epsilon}'}(\mathbf{r})
	\mathbf{E}^{(-)}_{\mathbf{k}',\boldsymbol{\epsilon}'}(\mathbf{r}')\ket{0}
	=\frac{-\hbar c }{\pi\epsilon_{0}}\int dk'\:k^{'2}
	\textrm{Im}\mathbb{G}(\mathbf{r}-\mathbf{r}';k').\nonumber
\end{equation}
Here,  $\mathbb{G} (\mathbf{r}-\mathbf{r}';k')$ is the dyadic Green's function of the electric field induced at $\mathbf{r}$ by a source of 
frequency $\omega'$ located at $\mathbf{r}'$, 
\begin{equation}
	\mathbb{G} (\mathbf{R};\omega')=-\frac{k' e^{ik'R}}{4\pi}\left[\frac{\mathbb{P}}{k'R}+\frac{i\mathbb{Q}}{(k'R)^{2}}
	-\frac{\mathbb{Q}}{(k'R)^{3}}\right],\nonumber
\end{equation}
where the tensors $\mathbb{P}$ and $\mathbb{Q}$ are $\mathbb{P}=\mathbb{I}-\mathbf{R}\mathbf{R}/R^{2}$,  $\mathbb{Q}=\mathbb{I}-3\mathbf{R}\mathbf{R}/R^{2}$, 
with $\mathbf{R}=\mathbf{r}-\mathbf{r}'$, $k'=\omega'/c$.

\subsection{Incoherent evolution of the system}
The formalism developed in Refs.\cite{Acevedo,myPRA} allows for the effective implementation of incoherent effects  within  $\mathbb{U}$ in a way compatible with reduced unitarity. This means that, except for the radiative processes associated to the spontaneous decay of the auxiliary excited state, inherent to the incoherent pump procedure, our approach captures all possible radiative processes involving the probe field and the atomic system. As a result, the coherent transitions from the steady to any intermediate state of the active atom get attenuated in time at an effective rate $\Gamma/2$, with those intermediate states being different to the initial ones in each process.  In contrast, as for the passive atom, the attenuating factor  $\gamma/2$ applies only to the spontaneous decay from the upper state $e$. 

Finally, some comments are in order in regard to the perturbative nature of our approach. In the first place, the fact that the probe field is resonant with the transition frequency, $|\omega-\omega_{0}|\ll\omega_{0}$, makes it possible to discard quantum processes in which the intermediate states are not quasi-degenerate, i.e., that do not possess an energy close to that of the initial states. 
Second, the weak character of the probe field implies $\Omega_{0}\ll\gamma_{0}$, which implies that we can restrict ourselves to processes in which only two probe field photons are annihilated at the atomic system. Third, in regard to the perturbative nature of the collective effects, this is guaranteed by the condition $\frac{k_0^2}{\hbar\epsilon_{0}}|\boldsymbol{\mu}\cdot\mathbb{G}(\mathbf{R};\omega)\cdot\boldsymbol{\mu}|\ll\gamma_{0}$, 
which implies that the transfer of the excitation between the atoms is slow enough to restrict ourselves to processes in which  only one virtual photon is exchanged between them. Note that, in contrast to the classical approach, in which effective polarizabilities can be derived applying linear response theory for any interatomic distance compatible with the electric dipole approximation --see Sec.\ref{lasec6}, no analogous treatment can be carried out in the quantum approach. The reason being the need to implement order by order the pump mechanism, in a way compatible with the preservation of unitarity.

\section{Scattered emission. General formulation}\label{lasec3}

\begin{figure}
		\begin{center}
			\includegraphics[width=140mm,angle=0,clip]{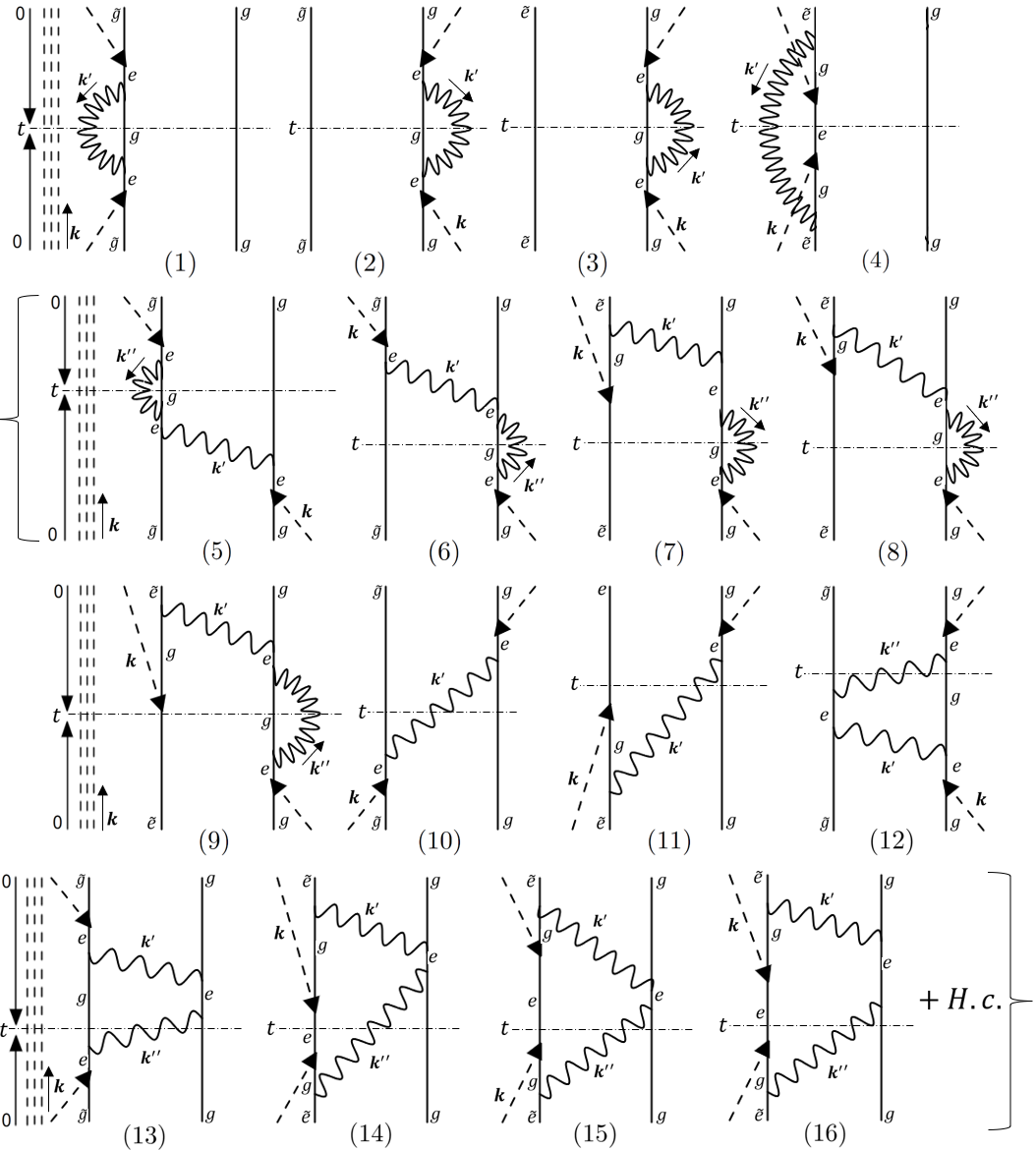}
			\caption{Diagrammatic representation of scattering processes at leading order. Time is represented along the vertical axis, extending from the initial time 0 to the observation time $t$. The upper amplitudes correspond to the evolution generated by $\mathbb{U}^{\dagger}(t)$, while the lower amplitudes correspond to the evolution generated by $\mathbb{U}(t)$. Solid straight lines denote the atomic states $g$ and  $e$. The tilded states, $\tilde{g}$ and $\tilde{e}$, appearing at the initial time, represent the pure-state components of the mixed state of atom $A$. Wavy lines denote virtual photons that are created and annihilated at interaction vertices associated with either atom. Dashed arrows represent photons from the external field that are annihilated at interaction vertices on either atom. Each interaction vertex corresponds to an insertion of the interaction Hamiltonian $W$. A horizontal dash line indicates the state of the system at the observation time $t$. Processes (1-4) correspond to single-atom scattering, while (5-16) and their hermitian conjugates (H.c) correspond to collective scattering.} \label{fig2}
		\end{center}
	\end{figure}

Following the classification of radiative processes outlined in Refs.\cite{Acevedo,myPRA}, we identify the scattering processes with those depicted diagrammatically in Fig.\ref{fig2}.  The first four correspond to single-atom scattering, from either atom in its ground state [diagrams (1, 2, 3)], and from the active atom in its excited state [(4)]. All the rest of the processes correspond to collective scattering and involve both atoms, the active one on the left, the passive one on the right in each diagram. 

More specifically, those diagrams   represent the probability of each scattering process, $P_{n}(t)=|\langle\Psi_{n}^{f}|\Psi_{n}(t)\rangle|^{2}$, $n=1,..,16$.  
In each diagram $n$ the initial state $|\Psi_{n}(0)\rangle$ is one of the pure steady states, $|\Psi_{0}\rangle_{g/e}$, which is evolved up to the observation time $t$ with a given element of the $\mathbb{U}$-matrix, $\mathbb{U}_{n}(t)$, such that $|\Psi_{n}(t)\rangle=\mathbb{U}_{n}(t)|\Psi_{n}(0)\rangle=|\Psi_{n}^{f}\rangle$, where the initial and final states defer only on one photon. That is, one of the probe photons in $|\Psi_{n}(0)\rangle$ is replaced in 
$|\Psi_{n}^{f}\rangle$ with a scattered photon of undefined frequency $\omega'$, momentum $\mathbf{k}'$, and polarization $\boldsymbol{\epsilon}'$. Since we are interested in the emission directionality, in contrast to the calculations performed in Ref.\cite{Acevedo}, only the sum over polarization states is carried out over the scattered photons. Specifically, the initial and final states of each scattering process of Fig.\ref{fig2} are
\begin{eqnarray}
	\fl	|\Psi_{m}(0)\rangle&=&|\Psi_{0}\rangle_{g},\:\:|\Psi_{m}^{f}\rangle=\sum_{\mathbf{k}',\boldsymbol{\epsilon}'}|(N-1)_{\mathbf{k},\boldsymbol{\epsilon}},1_{\mathbf{k}',\boldsymbol{\epsilon}'};g,g\rangle,\quad m=1,2,5,6,10,12,13;\nonumber\\
	\fl	|\Psi_{m}(0)\rangle&=&|\Psi_{0}\rangle_{e},\:\:|\Psi_{m}^{f}\rangle=\sum_{\mathbf{k}',\boldsymbol{\epsilon}'}|(N-1)_{\mathbf{k},\boldsymbol{\epsilon}},1_{\mathbf{k}',\boldsymbol{\epsilon}'};e,g\rangle,\quad m=3,4,7,8,9,11,14,15,16.\nonumber
\end{eqnarray}
The quantity of interest is the differential power emitted into a direction forming an angle $\theta$ with
the interatomic axis. The relevant contributions can be grouped according to whether the interfering
photons are emitted by the same atom or by different atoms. This classification is useful because the
two groups acquire different phase factors under reversal of the illumination direction. Thus, from the diagrams of Fig.\ref{fig2} we differentiate between two kinds of processes, namely, those in which the scattered power results from the interference of two photons emitted by the same atom (i.e., from self-interference), either $A$ [diagrams (1), (4) and (5) of Fig.\ref{fig2} and their hermitian conjugates (h.c.)] or $B$ [diagrams (2), (3), (6), (7), (8) and (9) and their h.c.]; and those in which the two interfering photons proceed from different atoms each [diagrams (10) to (16) of Fig.\ref{fig2} and their h.c.].  The scattering corresponding to the former will be referred to as each-atom scattering, whereas that corresponding to the latter will be referred to as both-atoms scattering. 

Further, for each-atom scattering we distinguish between single-atom scattering processes  [diagrams (1-4) of Fig.\ref{fig2}] and collective-scattering processes  [diagrams (5-9) of Fig.\ref{fig2} and h.c.]. In the former, the dipole moments at the emitting atom is induced solely by the probe field. In the latter, one of the dipole moments at the emitting dipole is induced by the probe field, while the other moment is induced by the field sourced by the non-emitting atom whose dipole moment itself is induced by the probe field. Hence, in diagrams (5-9) of Fig.\ref{fig2} the probe field strikes both atoms, and the photon flying between them carries the induction field.

Likewise, for both-atoms scattering we distinguish between processes which results from the interference between the amplitudes of single-scattering ones [diagrams (10) and (11) of Fig.\ref{fig2} and h.c.], and those which result from the interference between the amplitudes of single-scattering processes and each-atom collective scattering ones  [diagrams (12) to (16) of Fig.\ref{fig2} and h.c.]. In the former, the dipole moment on each emitting atom is induced by the probe field. For instance, diagram (11) results from the interference of diagrams (3) and (4). In the latter, only one of the dipole moments is induced by the probe field, whereas the dipole moment on the other atom is induced by the field sourced by the previous atom whose dipole moment itself is induced by the probe field. Hence, in diagrams (12-16) of Fig.\ref{fig2} the probe field strikes only one of the atoms, and the photon flying between them carries the induction field. For instance, in diagram (14), which results from the interference of diagrams (4) and (9), the probe field only strikes the active atom, and the photon flying between the atoms induces the dipole moment on the passive atom.
	
The quantity of our interest is the differential scattering power which, from each diagram, is calculated as the time derivative of the EM energy of its final state,
\begin{equation}
\fl	\mathcal{W}_{sc}(\theta)=\sum_{n=1}^{16}\frac{1}{\sin{\theta}}\frac{\mathop{d}^{2}}{\mathop{dtd\theta}}\langle\Psi_{n}(t)|\Psi_{n}^{f}\rangle\langle\Psi_{n}^{f}|H_{EM}|\Psi_{n}^{f}\rangle\langle\Psi_{n}^{f}|\Psi_{n}(t)\rangle,\quad\gamma t\gtrsim1, \label{powerscat} 
\end{equation}
where $\theta$ is the angle between the scattered radiation and the interatomic axis. 

Next, we will compute the power scattered by the system for two kinds of illumination. That is, for frontal illumination, in which the light incidence is orthogonal to the interatomic axis; and for side lighting, in which the incidence is along the interatomic axis, either from the active side or the passive side.

\section{Results: Scattered emission for frontal illumination}\label{lasec4}

For light incidence orthogonal to the interatomic axis, the probe field strikes both atoms with the same phase. Thus, no phase-shift factor appears in the calculation of any scattering process, irrespective of whether the external photons are absorbed by an only atom or both. We will distinguish between each-atom  scattered power, associated to diagrams $(1-9)$ of Fig.\ref{fig2}, and both-atoms scattered power associated to diagrams $(10-16)$. 

\subsection{Each-atom scattered radiation}

The processes related to each-atom scattering, diagrams $(1)$ to $(9)$ in Fig.\ref{fig2}, are those which result from the interference of two scattering amplitudes in which the scattered photons are emitted from the same atom.  In single-scattering processes, diagrams (1) to (4), only one of the atoms absorbs photons from the probe field and that atom emits independently, with no exchange of virtual photons with the other. In collective-scattering processes, diagrams $(5-9)$, both atoms absorb one  photon from the probe field and a virtual photon is exchanged between them, inducing a correlation between their dipole moments. The latter generates a factor  $\boldsymbol{\mu}\cdot\mathbb{G}(\omega ;\textbf{R})\cdot\boldsymbol{\mu}$ in their contribution to the scattering power. Following Ref.\cite{Acevedo}, we will find convenient to define 
\begin{equation}
	\fl\Omega(R)=\frac{k_0^2}{\hbar\epsilon_{0}} \boldsymbol{\mu}\cdot \left[\textrm{Re}\mathbb{G}(\mathbf{R};\omega) + i \textrm{Im}\mathbb{G}(\mathbf{R};\omega) \right]\cdot\boldsymbol{\mu}
	\equiv \widetilde{\Omega}(R)-i\widetilde{\Gamma}(R).
\end{equation}
For the sake of illustration, the calculation of the scattered power associated with diagram (6)  is given in detail in the Appendices.

Adding up the contribution from diagrams $(1)$ to $(9)$ we arrive at the following expression for the each-atom ($EA$) differential scattering power for frontal illumination ($\perp$), 
\begin{eqnarray} 
	\fl	\mathcal{W}_{\perp}^{EA}(\theta)
		 &= \frac{ \Omega_{0}^{2} \omega^{4} |\boldsymbol{\mu}|^{2}}{16 \pi \epsilon_{0} c^{3} } (1+\cos^{2}{\theta}) \left[\frac{1}{2\left[
			(\omega-\omega_{0})^{2}+\frac{\gamma^{2}}{4} \right]} + \frac{1}{2\left[(\omega-\omega_{0})^{2}+\frac{\Gamma^{2}}{4}\right]} \right.\label{sympowerscatperp}\\
		\fl	&+ \left.
		\frac{\gamma \left[2(\omega-\omega_{0}) \tilde{\Omega}(R)-\left(\frac{\gamma+\Gamma}{2}\right) \tilde{\Gamma}(R) \right] }{\Gamma \left[(\omega-\omega_{0})^{2}+\frac{\gamma^{2}}{4}\right] \left[(\omega-\omega_{0})^{2}+\frac{\Gamma^{2}}{4} \right]} \right.\nonumber\\
	\fl	& -\left.\frac{2 \mathcal{P} \left[[(\omega-\omega_{0})^{2}+\frac{\Gamma^{2}}{4}]\tilde{\Gamma}(R)-[(\omega-\omega_{0})^{2}-\frac{\gamma \Gamma}{4}]\tilde{\Gamma}(R)-(\frac{\gamma+\Gamma}{2}) (\omega-\omega_{0}) \tilde{\Omega}(R) \right]}{\Gamma(\gamma+\Gamma) [(\omega-\omega_{0})^{2}+\frac{\Gamma^{2}}{4}] [(\omega-\omega_{0})^{2}+\frac{\gamma^{2}}{4}]} \right],\nonumber
\end{eqnarray}
where the first two terms on the right hand side correspond to single-atom scattering [diagrams (1-4)], the third term does to diagrams (5) and (6), and the last term to diagrams (7) and (9). 
Note that the final expression remains invariant under parity, i.e., under the transformation $\theta\rightarrow\theta+\pi$. Hence, this is the parity-symmetric differential scattering power for frontal illumination. Nonetheless, this radiation is not isotropic since it depends on $\theta$. However, this is just due to the anisotropy of the transition dipole moment $\boldsymbol{\mu}$, which has been chosen  orthogonal to the interatomic axis. Interestingly, Eq.(\ref{sympowerscatperp}) contains terms proportional to  $(\omega-\omega_{0})\tilde{\Omega}(R)$ which makes the scattering spectrum asymmetric with respect to the resonant frequency. This term has its origin in the field carried by the flying photon of diagrams (5-9), which is sourced by the dipole of one the atoms and induces a dipole moment on the other atom.

\subsection{Both-atoms scattered radiation}

The processes associated with both-atoms scattering, diagrams $(10)$ to $(16)$ and their h.c. in Fig.\ref{fig2}, are those which result from the interference of two scattering amplitudes in which the scattered photons are emitted from different atoms. As explained in Sec.\ref{lasec3}, in diagrams (10) and (11) both atoms absorb one photon from the probe field, whereas in diagrams (12-16) only one the atoms absorbs photons from the probe field and sources the electric field which induces a dipole moment on the other atom. For the sake of illustration, the calculation of the scattered power associated with diagrams (11) and (14) of Fig.\ref{fig2} are given in detail in the Appendices.

Adding up the contribution from diagrams $(10)$ and $(11)$ and their h.c.,  we arrive at the following expression for the both-atoms (BA) differential scattering power for frontal illumination ($\perp$), 
\begin{eqnarray} 
		\fl	\mathcal{W}_{\perp}^{BA}(\theta)
		&=\frac{ \Omega_{0}^{2} \omega^{4} |\boldsymbol{\mu}|^{2}}{16 \pi \epsilon_{0} c^{3} } \; (1+\cos^{2}{\theta}) \;\Bigl\{ \frac{\gamma}{\;\Gamma} \frac{ \left[(\omega-\omega_{0})^{2} + \frac{\gamma \Gamma}{4}\right] \cos{(kR \cos{\theta})} - \left(\frac{\Gamma - \gamma}{2}\right) (\omega-\omega_{0}) \sin{(kR \cos{\theta})} }{\left[(\omega-\omega_{0})^{2}+\frac{\gamma^{2}}{4}\right] \left[(\omega-\omega_{0})^{2}+\frac{\Gamma^{2}}{4}\right]}\nonumber\\
	\fl	&-\frac{\mathcal{P}}{\;\Gamma}\frac{ \left[(\omega-\omega_{0})^{2} - \frac{\gamma \Gamma}{4}\right] \cos{(kR \cos{\theta})} + \left(\frac{\Gamma + \gamma}{2}\right) (\omega-\omega_{0}) \sin{(kR \cos{\theta})} }{\left[(\omega-\omega_{0})^{2}+\frac{\gamma^{2}}{4}\right] \left[(\omega-\omega_{0})^{2}+\frac{\Gamma^{2}}{4}\right]}\nonumber\\
	\fl	&-\;\frac{\gamma}{\;\Gamma} \frac{ \frac{(\Gamma+\gamma)}{2}\cos{(kR \cos{\theta})}\tilde{\Gamma}(R) -\left[2(\omega-\omega_{0})\cos{(kR \cos{\theta})}-
			\frac{(\Gamma-\gamma)}{2}\sin{(kR \cos{\theta})}\right]\tilde{\Omega}(R)}{\left[(\omega-\omega_{0})^{2}+\frac{\gamma^{2}}{4}\right] \left[(\omega-\omega_{0})^{2}+\frac{\Gamma^{2}}{4}\right]}\nonumber\\
	\fl	&+\frac{\mathcal{P}}{\;\Gamma} \; \frac{2[\sin{(kR\cos{\theta})} \tilde{\Omega}(R)- \cos{(kR \cos{\theta})}\tilde{\Gamma}(R)]}
		{(\gamma+\Gamma)\left[(\omega-\omega_{0})^{2}+\frac{\Gamma^{2}}{4}\right] }\nonumber\\
	\fl	&+\frac{\mathcal{P}}{\;\Gamma}
		\left[ \frac{2\left[(\omega-\omega_{0})\left(\frac{\gamma+\Gamma}{2}\right)\cos{(kR \cos{\theta})}-\left[(\omega-\omega_{0})^{2}-\frac{\gamma\Gamma}{4}\right]
			\sin{(kR \cos{\theta})}\right]\tilde{\Omega}(R)}{(\gamma+\Gamma)\left[(\omega-\omega_{0})^{2}+\frac{\gamma^{2}}{4}\right] \left[(\omega-\omega_{0})^{2}+\frac{\Gamma^{2}}{4}\right]} \right.\nonumber\\
	\fl	& + \left. \frac{2\left[\left[(\omega-\omega_{0})^{2}-\frac{\gamma\Gamma}{4}\right]
			\cos{(kR \cos{\theta})}+(\omega-\omega_{0})\left(\frac{\gamma+\Gamma}{2}\right)\sin{(kR \cos{\theta})}\right]\tilde{\Gamma}(R)}{(\gamma+\Gamma)\left[(\omega-\omega_{0})^{2}+\frac{\gamma^{2}}{4}\right] \left[(\omega-\omega_{0})^{2}+\frac{\Gamma^{2}}{4}\right]} \right]\Bigr\}, \label{asympowerscatperp} 
	\end{eqnarray}
where the first term on the right hand side of the equation corresponds to  diagram (10), the second term to  diagram (11), the third term to diagrams (12) and (13) combined, the fourth term to diagram (14) and the last one proportional to $\mathcal{P}/\Gamma$ to diagram (16), respectively. Those terms proportional to $\sin{(kR \cos{\theta})}$ are the ones which breaks parity, since they change sign under the transformation $\theta\rightarrow\theta+\pi$. Hence, this is the parity-asymmetric differential scattering power for frontal illumination. 

\begin{figure}
	\begin{center}
		\includegraphics[width=110mm,angle=0,clip]{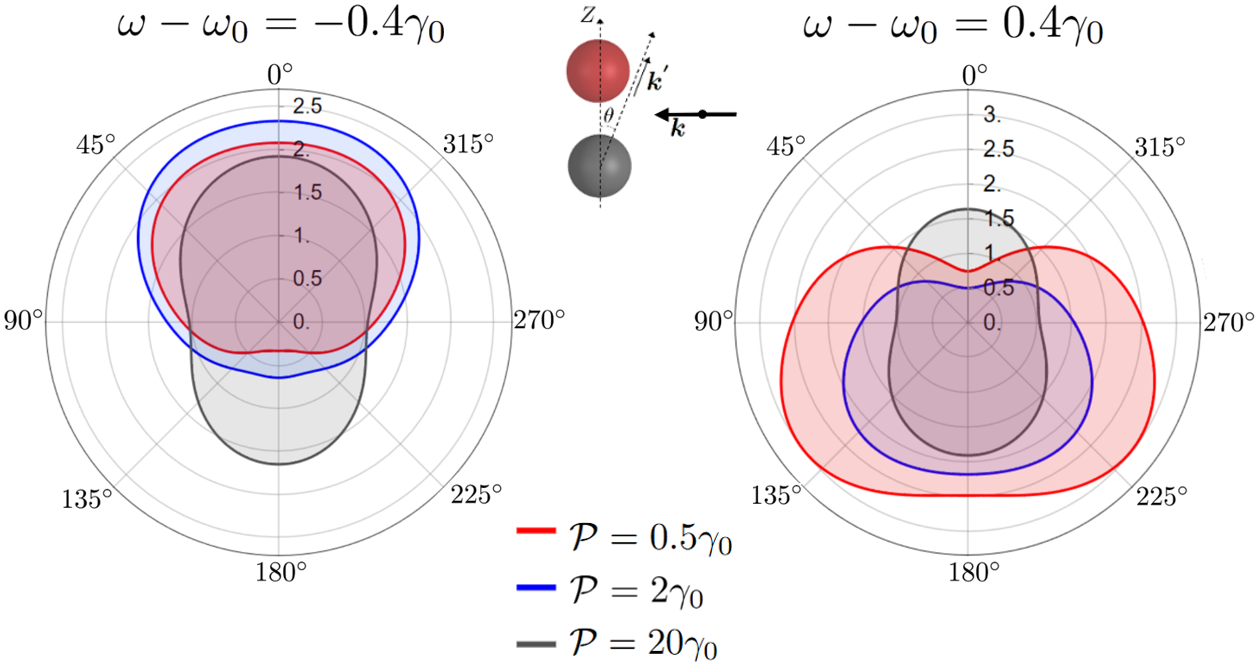}
		\caption{Angular distribution of the scattered power for frontal incidence $\mathcal{W}_{\perp}(\theta)$ and different pump rates, $\mathcal{P}=(0.5,2,20)\gamma_{0}$. The interatomic distance is taken at $kR=2$, the non-radiative linewidth is $\gamma_{nr}=0.2\gamma_{0}$, and the detuning is fixed at values $\omega-\omega_{0}=-0.4\gamma_{0}$ (left panel) and $\omega-\omega_{0}=0.4\gamma_{0}$ (right panel).} \label{fig5}
	\end{center}
\end{figure}

\begin{figure}
	\begin{center}
		\includegraphics[width=130mm,angle=0,clip]{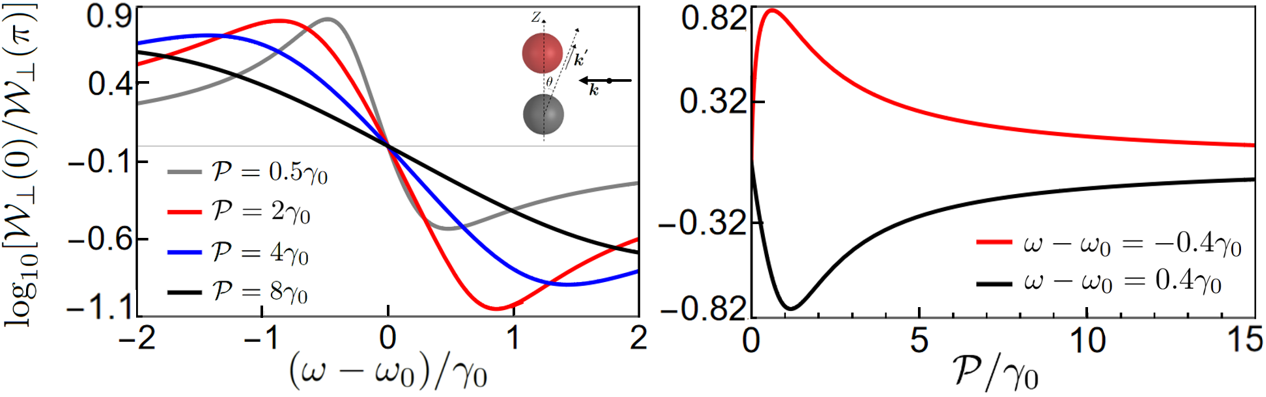}
		\caption{Graphical representation of  the log of the ratio of the scattered power towards the active atom and towards the passive one for frontal incidence, $\log_{10}{[\mathcal{W}_{\perp}(0)/\mathcal{W}_{\perp}(\pi)]}$, for fixed interatomic distance $kR=2$, and non-radiative linewidth $\gamma_{nr}=0.2\gamma_{0}$. Left panel: In terms of the detuning $\omega-\omega_{0}$, with several pump rates, $\mathcal{P}=(0.5,2,4,8)\gamma_{0}$. Right panel:  In terms of the pump rate $\mathcal{P}$, with detuning  $\omega-\omega_{0}=0.4\gamma_{0}$ (black line) and $\omega-\omega_{0}=-0.4\gamma_{0}$ (red line).} \label{fig3}
	\end{center}
\end{figure}

We analyse in Fig.\ref{fig5} the angular distribution of the scattered power for different pump rates of the active atom, $\mathcal{W}_{\perp}=\mathcal{W}^{EA}_{\perp}+\mathcal{W}^{BA}_{\perp}$. In the left panel of Fig.\ref{fig3} we represent the ratio between the scattered power towards the active atom ($\theta=0$) and the passive one ($\theta=\pi$), for the interatomic distance $kR=2$, $\gamma_{nr}=0.2\gamma_{0}$, and different pump rates. We observe that the scattered power is greater towards the active atom for negative values of the detuning, greater towards the passive one for positive detuning, and still a small asymmetry exists at exact resonance. We note, however, that all the asymmetric terms are proportional to $\pm\sin{(kR)}$, and thus, its sign depends on the interatomic distance. In the right panel of Fig.\ref{fig3} we observe that the asymmetry reduces as the pump rate increases. 
In particular, the asymmetry with respect to exact resonance increases with  the pump rate until $\mathcal{P}\approx2\gamma_{0}$ and the symmetry gets restored for $\mathcal{P}\ll\gamma_{0}$ and $\mathcal{P}\gtrapprox10\gamma_{0}$.

\section{Results: Scattered emission for side lighting}\label{lasec5}

For light incidence along to the interatomic axis, we will distinguish between scattered power for incidence from the active atom, i.e., from the gain side (G); and for incidence from the passive atom, i.e., from the loss side (L). In contrast to frontal illumination, where parity-asymmetry was found in the difference of the radiation emitted towards the active or the passive atom, here we will be interested in the asymmetry of the forward and backward scattering radiation with respect to the side of incidence of the probe field. In this case, since the external field strikes each atom  with different phases, those processes in which both atoms absorb a probe field photon each, will carry a phase-shift factor $\exp{(\pm ikR)}$ for incidence from either side. Along the same lines as for frontal illumination, we will differentiate between each-atom scattering and both-atoms scattering. We will see that, in contrast to frontal illumination, the contribution of the former to forward scattering does depend on the side of incidence, and so does part of the contribution of the latter. 

\subsection{Each-atom scattered radiation}

In this case, we differentiate between  single-atom scattering and collective each-atom scattering. It is in the latter that an asymmetrical contribution will be found with respect to the side of incidence, as it contains processes in which the probe field impinges each atom with different phases. 

For the sake of illustration, the calculations of the scattered power associated with diagram (6), for both incidence from the gain side (G) and from the loss side (L)  are given in detail in the Appendices.

Adding up the contribution from diagrams $(1)$ to $(9)$ for both sides of incidence, gain and loss, we arrive at the following expression for the each-atom ($EA$) differential scattering power for side lighting, 
	\begin{eqnarray} 
		\fl	&\mathcal{W}_{G,L}^{EA}(\theta)
			 = \frac{ \Omega_{0}^{2} \omega^{4} |\boldsymbol{\mu}|^{2}}{16 \pi \epsilon_{0} c^{3} } (1+\cos^{2}{\theta}) \left[\frac{1}{2\left[(\omega-\omega_{0})^{2}+\frac{\gamma^{2}}{4} \right]} + \frac{1}{2\left[(\omega-\omega_{0})^{2}+\frac{\Gamma^{2}}{4}\right]} \right.\nonumber\\
		\fl	& + \left. 2\frac{\mathcal{P}}{\Gamma} \Bigl[\frac{\left[\tilde{\Gamma}(R)\cos{kR}\pm\tilde{\Omega}(R)\sin{kR}\right] \left[(\omega-\omega_{0})^{2}-\frac{\gamma\Gamma}{4}\right]}{(\gamma+\Gamma) [(\omega-\omega_{0})^{2}+\frac{\Gamma^{2}}{4}] [(\omega-\omega_{0})^{2}+\frac{\gamma^{2}}{4}]}\right.\nonumber\\
		\fl &- \left. \frac{\left[\tilde{\Omega}(R)\cos{kR}\mp\tilde{\Gamma}(R) \sin{kR}\right] \left(\frac{\gamma+\Gamma}{2}\right)(\omega-\omega_{0})}{(\gamma+\Gamma) [(\omega-\omega_{0})^{2}+\frac{\Gamma^{2}}{4}] [(\omega-\omega_{0})^{2}+\frac{\gamma^{2}}{4}]}- \frac{\left[\tilde{\Gamma}(R) \cos{kR}\pm\tilde{\Omega}(R)\sin{kR} \right] }{(\gamma+\Gamma)\left[(\omega-\omega_{0})^{2}+\frac{\gamma^{2}}{4}\right]}\Bigr]\right.\nonumber\\
		\fl	& + \left. \frac{\gamma}{\Gamma} \frac{\tilde{\Omega}(R)\left[2(\omega-\omega_{0})\cos{kR}\mp\left(\frac{\Gamma-\gamma}{2}\right)\sin{kR}\right]-\tilde{\Gamma}(R)\left(\frac{\gamma+\Gamma}{2}\right)\cos{kR}}{\left[(\omega-\omega_{0})^{2}+\frac{\gamma^{2}}{4}\right] \left[(\omega-\omega_{0})^{2}+\frac{\Gamma^{2}}{4} \right]}\right].\label{WGLea}
	\end{eqnarray}
In this expression, the first two terms within the square brackets correspond to single-atom scattering and are common to the scattering for both sides of incidence. In the rest of the terms, corresponding to collective scattering, those terms proportional to $\tilde{\Omega}(R)\sin{kR}$ and $\tilde{\Gamma}(R)\sin{kR}$ enter with opposite sign in the scattering for each side of incidence. They are responsible of the asymmetry of scattering in any direction. Note that, in contrast to the case of frontal illumination, this asymmetry results from the phase difference between the probe field photons that  strike each atom. Therefore, we interpret that this asymmetry results from the interference between the two fields emitted by the same atom, one of them sourced by its dipole moment induced by the probe field, and the other one sourced by its dipole moment induced by the field generated by the other atom.

\subsection{Both-atoms scattered radiation}

In this case, the scattered power proceeds from the interference of two photons emitted from different atoms. However, while in diagrams (10) and (11) the dipole moments on each atom are sourced by the probe field, in diagrams (12-16) only the dipole moment of one of the atoms is induced by the probe field. Thus, the contribution of the processes of diagrams $(10)$ and $(11)$ and their h.c. contain factors which depend on the phase difference $\pm kR$ accumulated by the probe field in its way between both atoms, while those factors do not appear in diagrams (12-16) where the probe field strikes only one of the atoms. Interestingly, for forward scattering, in diagrams  $(10)$ and $(11)$  the interference phase between the emitted photons is compensated with the phase difference between the probe fields that strike each atom. However, this cannot be the case for the rest of the processes. Adding up all the contributions for both sides of incidence, gain $(G)$ and loss $(L)$, we arrive at the following expression for the both-atoms ($BA$) differential scattering power for side lighting,
	\begin{eqnarray}
	\fl		\mathcal{W}_{G,L}^{BA}(\theta)
			&=\frac{ \Omega_{0}^{2} \omega^{4} |\mu|^{2}}{16 \pi \epsilon_{0} c^{3} } \; (1+\cos^{2}{\theta}) \; \nonumber\\
	\fl		& \times \Bigl\{\frac{\gamma}{\;\Gamma}\frac{\left[(\omega-\omega_{0})^{2} + \frac{\gamma \Gamma}{4}\right] \cos{[kR(\cos{\theta}\pm1)]} - 
				\left(\frac{\Gamma - \gamma}{2}\right) (\omega-\omega_{0}) \sin{[kR(\cos{\theta}\pm1)]}}{\left[(\omega-\omega_{0})^{2}+\frac{\gamma^{2}}{4}\right] \left[(\omega-\omega_{0})^{2}+\frac{\Gamma^{2}}{4}\right]}\nonumber\\
	\fl		& -\frac{\mathcal{P}}{\;\Gamma}\frac{\left[(\omega-\omega_{0})^{2} - \frac{\gamma \Gamma}{4}\right] \cos{[kR(\cos{\theta}\pm1)]} + 
				\left(\frac{\Gamma + \gamma}{2}\right) (\omega-\omega_{0}) \sin{[kR(\cos{\theta}\pm1)]}}{\left[(\omega-\omega_{0})^{2}+\frac{\gamma^{2}}{4}\right] \left[(\omega-\omega_{0})^{2}+\frac{\Gamma^{2}}{4}\right]}\nonumber\\
	\fl		&+\frac{\gamma}{\;\Gamma} \frac{ \left[2(\omega-\omega_{0})\cos{(kR \cos{\theta})}-
				\frac{(\Gamma-\gamma)}{2}\sin{(kR \cos{\theta})}\right]\tilde{\Omega}(R)  - \frac{(\Gamma+\gamma)}{2}\cos{(kR \cos{\theta})}\tilde{\Gamma}(R) }{\left[(\omega-\omega_{0})^{2}+\frac{\gamma^{2}}{4}\right] \left[(\omega-\omega_{0})^{2}+\frac{\Gamma^{2}}{4}\right]}\nonumber\\
	\fl	&+\frac{\mathcal{P}}{\;\Gamma} \; \frac{2[\sin{(kR\cos{\theta})} \tilde{\Omega}(R)- \cos{(kR \cos{\theta})}\tilde{\Gamma}(R)]}
			{(\gamma+\Gamma)\left[(\omega-\omega_{0})^{2}+\frac{\Gamma^{2}}{4}\right] }\nonumber\\
		\fl	&+\frac{\mathcal{P}}{\;\Gamma}
			\left[ \frac{2\left[(\omega-\omega_{0})\left(\frac{\gamma+\Gamma}{2}\right)\cos{(kR \cos{\theta})}-\left[(\omega-\omega_{0})^{2}-\frac{\gamma\Gamma}{4}\right]
				\sin{(kR \cos{\theta})}\right]\tilde{\Omega}(R)}{(\gamma+\Gamma)\left[(\omega-\omega_{0})^{2}+\frac{\gamma^{2}}{4}\right] \left[(\omega-\omega_{0})^{2}+\frac{\Gamma^{2}}{4}\right]} \right.\nonumber\\
		\fl	& + \left. \frac{2\left[\left[(\omega-\omega_{0})^{2}-\frac{\gamma\Gamma}{4}\right]
				\cos{(kR \cos{\theta})}+(\omega-\omega_{0})\left(\frac{\gamma+\Gamma}{2}\right)\sin{(kR \cos{\theta})}\right]\tilde{\Gamma}(R)}{(\gamma+\Gamma)\left[(\omega-\omega_{0})^{2}+\frac{\gamma^{2}}{4}\right] \left[(\omega-\omega_{0})^{2}+\frac{\Gamma^{2}}{4}\right]} \right]\Bigr\}.\label{WGLba}
	\end{eqnarray}
The first two terms on the r.h.s. of this equation corresponds to diagrams (10) and (11). The phase difference due to lighting from opposite sides corresponds to the $\pm kR$ term within the argument of the harmonic functions, which cancels the interference phase $kR\cos{\theta}$ for forward scattering. Note that forward ($fw$) scattering for incidence from the gain side implies $\theta_{G}^{fw}=\pi$, whereas forward scattering for incidence from the loss side implies $\theta_{L}^{fw}=0$. For these angles the argument of the harmonic functions vanishes in both cases, implying the invariance of diagrams (10) and (11) for the forward direction with respect to the side of incidence. In contrast, for backward scattering ($bw$) the two phase differences add up with opposite signs, thus enhancing the asymmetry.  In the rest of the terms [diagrams (12-16)] only the interference phase appears in the argument of the harmonic functions, which are common to both sides of incidence and are accompanied by factors $\tilde{\Omega}(R)$ and $\tilde{\Gamma}(R)$ that proceed from the EM field of the photons flying between the atoms.  For the sake of illustration, the calculation of the scattered power associated with diagrams (11) and (14), for incidence from both G and L, is given in detail in the Appendices.   

\subsection{Forward and backward scattering}

Finally, evaluating each-atom and both-atoms scattering powers for incidence from the gain and loss sides,
\begin{eqnarray}
	&\mathcal{W}_{G,L}(\theta)=\mathcal{W}_{G,L}^{EA}+\mathcal{W}_{G,L}^{BA},\textrm{ with }\mathcal{W}^{fw}_{G}=\mathcal{W}_{G}(\pi),\nonumber\\
	&\mathcal{W}^{fw}_{L}=\mathcal{W}_{L}(0),\:\mathcal{W}^{bw}_{G}=\mathcal{W}_{G}(0),\:\mathcal{W}^{bw}_{L}=\mathcal{W}_{L}(\pi).
\end{eqnarray}
From Eqs.(\ref{WGLea}) and (\ref{WGLba})  we verify that only those terms which depend on the phase difference between the probe fields impinging on each atom, $\propto\sin{(\pm kR)}$, those terms which depend on the interference phase between the photons emitted from each atom, $\propto\sin{(kR\cos\theta)}$, and those which depend on both, $\propto\sin{[kR(\cos\theta\pm1)]}$, may provide some asymmetry under the exchange $G\leftrightarrow L$ in forward and backward scattering. 
More specifically, those of the first kind correspond to diagrams (5-9), and affect in the same manner forward and backward scattering; those of the second kind correspond to diagrams (12-16), and contribute with opposite signs to forward and backward scattering; an those of the last kind, diagrams (10) and (11) only generate asymmetry in backward scattering. In addition, some of the asymmetric terms provided by diagrams (5-9) are compensated by those provided by diagrams (12-16). In particular, for forward scattering, the asymmetric terms of diagram (5) cancel with those of diagram (13); those of diagram (6) cancel with those of diagram (12). On the contrary, the asymmetric term of diagram (7) adds up to that of diagram (14); and only one of the asymmetric terms of diagram (9) cancels with one of those of diagram (16) while the other asymmetric terms add up to each other. Quite the reverse applies to backward scattering.

\begin{figure}
	\begin{center}
		\includegraphics[width=110mm,angle=0,clip]{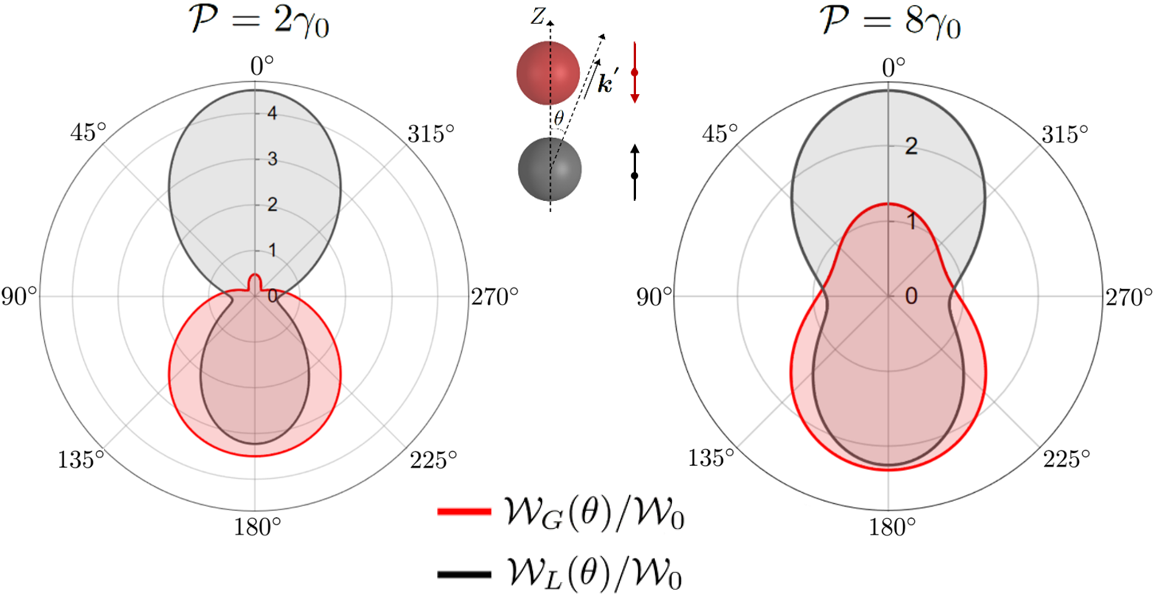}
		\caption{Angular distribution of the scattered power for side lighting, G in red, L in grey, and two different pump rates, $\mathcal{P}=2\gamma_{0}$ (left panel),  $\mathcal{P}=8\gamma_{0}$ (right panel). The interatomic distance is taken at $kR=2$, the non-radiative linewidth is $\gamma_{nr}=0.2\gamma_{0}$, and the detuning is fixed at the negative value $\omega-\omega_{0}=-0.4\gamma_{0}$.} \label{fig7}
	\end{center}
\end{figure}

\begin{figure}
		\begin{center}
			\includegraphics[width=145mm,angle=0,clip]{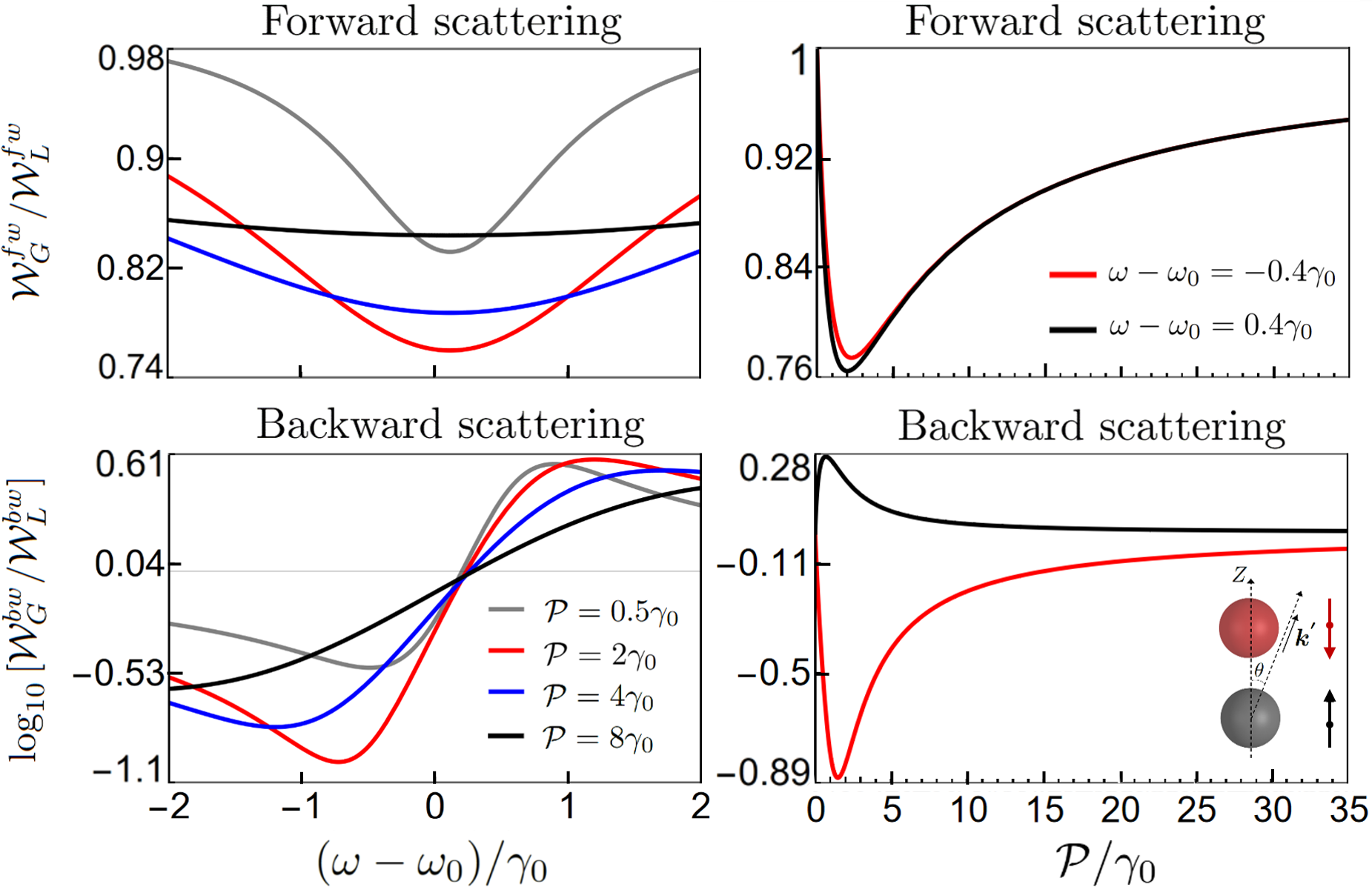}
			\caption{Upper left panel: graphical representation of the ratio of the forward-scattered powers for side illumination from the gain and loss sides, $\mathcal{W}_{G}^{fw}/\mathcal{W}_{L}^{fw}$, in terms of the detuning $\omega-\omega_{0}$. Lower left panel: graphical representation of the logarithm of the ratio between the backward-scattered powers for side illumination from the gain and loss sides, $\log_{10}{[\mathcal{W}_{G}^{bw}/\mathcal{W}_{L}^{bw}]}$, in terms of the detuning $\omega-\omega_{0}$.  The interatomic distance is taken at $kR=2$, the non-radiative linewidth is $\gamma_{nr}=0.2\gamma_{0}$, and several pump rates are considered, $\mathcal{P}=(0.5,2,4,8)\gamma_{0}$. Right panels: \emph{Idem} in terms of the detuning $\mathcal{P}$, for fixed detuning, $\omega-\omega_{0}=-0.4\gamma_{0}$ (red line), $\omega-\omega_{0}=0.4\gamma_{0}$ (black line).} \label{fig8}
		\end{center}
	\end{figure}

We analyse in Fig.\ref{fig7} the angular distribution of the scattered power for different pump rates of the active atom. In Fig.\ref{fig8}, upper left panel, we represent the ratio of the forward-scattered powers for side lighting from the gain and loss sides, $\mathcal{W}_{G}^{fw}/\mathcal{W}_{L}^{fw}$, for different pump rates. Likewise in  the lower left panel for the log of the ratio between the backward-scattered powers, $\log_{10}{[\mathcal{W}_{G}^{bw}/\mathcal{W}_{L}^{bw}]}$.
Interestingly, forward scattered power is greater for lighting from the loss side, for any pump rate and any detuning, whereas  backward scattered power is greater for G-lighting for negative values of the detuning, greater for L-lighting for positive detuning, and still some asymmetry exists at exact resonance. 
Note, however, that all the asymmetric terms are proportional to $\pm\sin{(kR)}$, and some of them are linear in $(\omega-\omega_{0})$.  Thus, the sign of the asymmetry depends on the interatomic distance as well as on the sign of the detuning. From the graphs of the right panels in Fig.\ref{fig8} we observe that the asymmetry increases with  the pump rate until $\mathcal{P}\approx2\gamma_{0}$ in both cases, and the symmetry gets restored for $\mathcal{P}\ll\gamma_{0}$ and $\mathcal{P}\gtrapprox40\gamma_{0}$.

\begin{figure}
	\begin{center}
		\includegraphics[width=135mm,angle=0,clip]{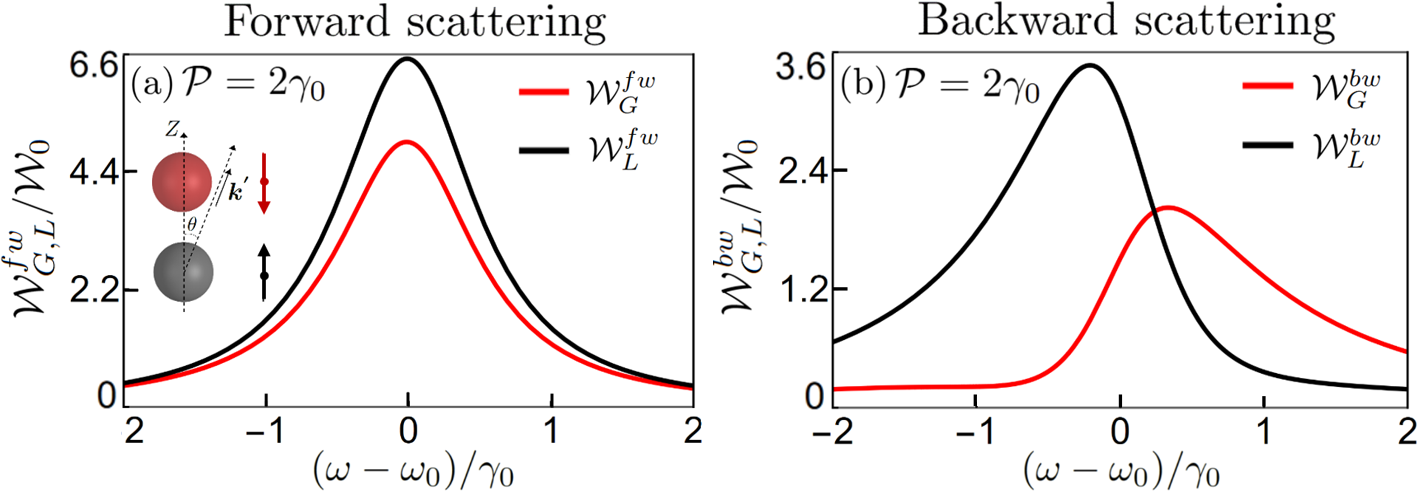}
		\caption{Graphical representation of the forward (left panel) and backward (right panel) scattered power for G-lighting (red line) and L-lighting (black line), normalized by $\mathcal{W}_{0}=\frac{ \Omega_{0}^{2} \omega^{4} |\boldsymbol{\mu}|^{2}}{16 \pi \epsilon_{0} c^{3} \gamma_{0}^{2}}$, in terms of the detuning $\omega-\omega_{0}$. The pump rate is $\mathcal{P}=2\gamma_{0}$, the interatomic distance is taken at $kR=2$ and the non-radiative linewidth is $\gamma_{nr}=0.2\gamma_{0}$.} \label{fig10}
	\end{center}
\end{figure}
In Fig.\ref{fig10}, left panel, we compare the forward scattered power for G-lighting with that for L-lighting in terms of the detuning. Likewise in the right panel for backward scattering. We observe that the difference is much more pronounced  for backscattering, and does not vanish at exact resonance. 

\section{Comparison with a classical approach}\label{lasec6}

In this Section we address the asymmetry of scattering within the framework of classical linear response theory. 

Let us take two  polarizable point particles sited at locations $\mathbf{R}_{A}$ and $\mathbf{R}_{B}$, with polarizabilities $\alpha_{A}$ and $\alpha_{B}$, and illuminated by an external probe field of frequency $\omega$. Considering as for the atoms that the induced dipole moments are orthogonal to the interparticle axis, and integrating over the azimuth angle $\phi$, the total scattered power in terms of the polar angle $\theta$ with respect to the interparticle axis reads \cite{Jacksonbook}
\begin{equation}
\fl	\mathcal{W}^{cl}(\theta)=\frac{\omega^{4}}{32\pi\epsilon_{0}c^{3}} \; (1+\cos^{2}{\theta})\;\left[  \; |\boldsymbol{p}_{A}|^{2} + |\boldsymbol{p}_{B}|^{2}
	+2\textrm{Re}\left\{ (\boldsymbol{p}_{A} \cdot \boldsymbol{p}^{*}_{B})  \; e^{-ikR\cos{\theta}} \right\} \right],\label{Wcl}
\end{equation}
where  $\boldsymbol{p}_{j}$  is the complex-valued electric dipole moment of each particle that oscillates at frequency $\omega$, with $j=A,B$.
The terms in Eq.(\ref{Wcl}) proportional to $ |\boldsymbol{p}_{A}|^{2} + |\boldsymbol{p}_{B}|^{2}$ are the classical analogs of the quantum each-atom scattering terms, whereas the term $2\textrm{Re}\left\{ (\boldsymbol{p}_{A} \cdot \boldsymbol{p}^{*}_{B})  \; e^{-ikR\cos{\theta}} \right\}$ is the classical analog of the quantum both-atoms scattering ones.

The problem of scattering for frontal and side lighting reduces now to one of computing the dipole moments induced on each particle in terms of the external electric field and the classical polarizabilities. To this end, we apply the classical equations of linear response theory and calculate in each case the effective polarizabilities of the particles which contain the mutual interaction between them and, for side lighting, the phase-shift between the fields incident on each particle.

Generically, the electric field induced by the particle $j$ at the point where particle $i$ is located reads
\begin{equation}
	\mathbf{E}(\omega;\mathbf{R}_{i})=-k^{2}\epsilon_{0}^{-1}\mathbb{G}(\omega;\mathbf{R}_{i},\mathbf{R}_{j})\cdot\mathbf{p}_{j},\quad i,j=A,B,\nonumber
\end{equation}
where it is implicit that all the functionals above, $\mathbf{E}$, $\mathbb{G}$, $\mathbf{p}_{j}$, are evaluated at the same frequency $\omega$. Hereafter we will denote these functionals evaluated at or between the active and passive particles with the subscripts $A$, $B$, respectively. Reciprocally, the dipole induced on each particle is the optical response of each one to the total field incident at its location,
\begin{equation}
	\boldsymbol{p}_{A}=\alpha_{A}[\boldsymbol{E}_{0,A}-\frac{k^{2}}{\epsilon_{0}}\mathbb{G}_{AB}\;\boldsymbol{p}_{B}],\quad
	\boldsymbol{p}_{B}=\alpha_{B}[\boldsymbol{E}_{0,B}-\frac{k^{2}}{\epsilon_{0}}\mathbb{G}_{BA}\;\boldsymbol{p}_{A}].\label{pg2}
\end{equation}
In these equations $\boldsymbol{E}_{0,A/B}$ are the probe fields of frequency $\omega$ incident at $\mathbf{R}_{A}$ and $\mathbf{R}_{B}$, respectively, and $\alpha_{A,B}$ are the electric polarizabilities of each atom as isolated in free space.

From these equations we realize that, generally, the induced dipoles cannot be written in terms of the probe field incident at each one solely, as they depend on the probe fields incident at both of them. However, for the cases of our interest, there exist a simple relationship between $\boldsymbol{E}_{0,B}$ and $\boldsymbol{E}_{0,A}$ that allows us to write  $\boldsymbol{p}_{A}=\boldsymbol{\alpha}_{A}\boldsymbol{E}_{0,A}$ and $\boldsymbol{p}_{B}=\boldsymbol{\alpha}_{B}\boldsymbol{E}_{0,B}$ for both frontal and side lighting.

\subsection{Frontal illumination}

For frontal illumination $\boldsymbol{E}_{0,A}=\boldsymbol{E}_{0,B}\equiv\boldsymbol{E}_{0}$, and Eq.(\ref{pg2}) 
can be written as
\begin{equation}
	\boldsymbol{p}_{A}=\boldsymbol{\alpha}^{\perp}_{A}\boldsymbol{E}_{0},\quad\boldsymbol{p}_{B}=\boldsymbol{\alpha}^{\perp}_{B}\boldsymbol{E}_{0}, \textrm{ with}\label{ps}
\end{equation}
\begin{eqnarray}
	\boldsymbol{\alpha}^{\perp}_{A}&=\alpha_{A}\left[\mathbb{I}-\frac{k^{2}}{\epsilon_{0}} \mathbb{G}_{AB}\alpha_{B}\right]\cdot\left[\mathbb{I}-\frac{k^{4}}{\epsilon_{0}^{2}}\alpha_{A}\mathbb{G}_{AB}\alpha_{B}\mathbb{G}_{BA}\right]^{-1},\label{alphaGtt}\\
	\boldsymbol{\alpha}^{\perp}_{B}&=\alpha_{B}\left[\mathbb{I}-\frac{k^{2}}{\epsilon_{0}} \mathbb{G}_{BA}\alpha_{A}\right]\cdot\left[\mathbb{I}-\frac{k^{4}}{\epsilon_{0}^{2}}\alpha_{B}\mathbb{G}_{BA}\alpha_{A}\mathbb{G}_{AB}\right]^{-1}.\label{alphaLtt}
\end{eqnarray}
Finally, replacing the expression of $\boldsymbol{p}_{A}$ and $\boldsymbol{p}_{B}$ of Eqs.(\ref{ps})-(\ref{alphaLtt}) in Eq.(\ref{Wcl}), we arrive at
\begin{equation}
\fl	\mathcal{W}^{cl}_{\perp}=\frac{\omega^{4}}{32\epsilon_{0}\pi c^{3}} \; (1+\cos^{2}{\theta})\;\bigl[  \; |\boldsymbol{\alpha}^{\perp}_{A}|^{2} + |\boldsymbol{\alpha}^{\perp}_{B}|^{2}
	+2\textrm{Re}\left\{ \boldsymbol{\alpha}^{\perp}_{A}\boldsymbol{\alpha}^{\perp*}_{B} \; e^{-ikR\cos{\theta}} \right\}\bigr]|\boldsymbol{E}_{0}|^{2}.\label{wclperp}
\end{equation}
Note that, comparing Eqs.(\ref{alphaGtt}) and (\ref{alphaLtt}), and the expression for $\mathcal{W}^{cl}_{\perp}$, we observe that asymmetry in scattering may be caused only by the interference between the fields emitted from each particle through the interference factor $ e^{-ikR\cos{\theta}}$ in Eq.(\ref{wclperp}), analogously to the quantum-atomic case --see Sec.\ref{lasec4}.

\subsection{Side lighting}

For side lighting we may write without loss of generality $\boldsymbol{E}_{0,B}=e^{i\mathbf{k}\cdot(\mathbf{R}_{B}-\mathbf{R}_{A})}\boldsymbol{E}_{0,A}$. Therefore, for lighting from the gain and from the loss side we may write $\boldsymbol{E}_{0,B}=e^{\pm i\:kR}\mathbf{E}_{0,A}$, respectively, and the induced dipole moments become
\begin{equation}
	\boldsymbol{p}_{A}=\boldsymbol{\alpha}^{G/L}_{A}\boldsymbol{E}_{0,A},\qquad\boldsymbol{p}_{B}=\boldsymbol{\alpha}^{G/L}_{B}\boldsymbol{E}_{0,B},\quad \textrm{ with}\label{psp}
\end{equation}
\begin{eqnarray}
	\boldsymbol{\alpha}^{G/L}_{A}&=\alpha_{A}\left[\mathbb{I}-\frac{k^{2}}{\epsilon_{0}}e^{\pm i\:kR}\mathbb{G}_{AB}\alpha_{B}\right]
	\cdot\left[\mathbb{I}-\frac{k^{4}}{\epsilon_{0}^{2}}\alpha_{A}\mathbb{G}_{AB}\alpha_{B}\mathbb{G}_{BA}\right]^{-1},\label{alphaGttp}\\
	\boldsymbol{\alpha}^{G/L}_{B}&=\alpha_{B}\left[\mathbb{I}-\frac{k^{2}}{\epsilon_{0}}e^{\mp i\:kR}\mathbb{G}_{BA}\alpha_{A}\right]\cdot\left[\mathbb{I}-\frac{k^{4}}{\epsilon_{0}^{2}}\alpha_{B}\mathbb{G}_{BA}\alpha_{A}\mathbb{G}_{AB}\right]^{-1},\label{alphaLttp}
\end{eqnarray}
where the $\pm$ or $\mp$ signs apply to lighting from the gain side and loss side, respectively. Taking $\boldsymbol{E}_{0,A}=\boldsymbol{E}_{0}$, and replacing the expression of $\boldsymbol{p}_{A}$ and $\boldsymbol{p}_{B}$ of Eqs.(\ref{psp})-(\ref{alphaLttp}) in Eq.(\ref{Wcl}), for both G and L lighting, we arrive at
\begin{equation}
	\fl\mathcal{W}_{G/L}^{cl}=\frac{\omega^{4}}{32\pi\epsilon_{0} c^{3}} \;(1+\cos^{2}{\theta})\;\bigl[\;|\boldsymbol{\alpha}^{G/L}_{A}|^{2} + |\boldsymbol{\alpha}^{G/L}_{B}|^{2}+2\textrm{Re}\left\{ \boldsymbol{\alpha}^{G/L}_{A}\boldsymbol{\alpha}^{G/L*}_{B} \; e^{\mp i\:kR}e^{-ikR\cos{\theta}} \right\}\bigr]|\boldsymbol{E}_{0}|^{2}.\label{Wclside}
\end{equation}
Interestingly, as found numerically in Ref.\cite{ManjavacasPT1}, we verify from Eq.(\ref{Wclside}) together with Eqs.(\ref{alphaGttp}), (\ref{alphaLttp})  that the power scattered in the forward direction is equal for both sides of illumination, whereas there exists an asymmetry for backward scattering. This is so despite the fact that the terms $|\boldsymbol{\alpha}^{G/L}_{A}|^{2} + |\boldsymbol{\alpha}^{G/L}_{B}|^{2}$ and $2\textrm{Re}\left\{ \boldsymbol{\alpha}^{G/L}_{A}\boldsymbol{\alpha}^{G/L*}_{B} \right\}$ in Eq.(\ref{Wclside}) contain asymmetric factors in them proportional to $\sin{kR}$. In particular, all each-atom scattering terms of the kind $|\alpha_{A}|^{2}\textrm{Re}\left\{ e^{\pm ikR}\mathbb{G}_{AB}\alpha_{B}\right\}$ and $|\alpha_{B}|^{2}\textrm{Re}\left\{ e^{\mp ikR}\mathbb{G}_{BA}\alpha_{A}\right\}$ cancel out with both-atoms scattering terms of the same kind. This is in contrast to the quantum-atomic case. 
Besides, as for the quantum case, the asymmetric factor $\sin{[kR(\mp1-\cos{\theta})]}$ in the both-atoms scattering term of Eq.(\ref{Wclside}) cancels out for forward scattering. As a result, classical forward scattering is invariant  under the exchange $G\leftrightarrow L$, which is a manifestation of reciprocity in the optical response of a linear classical system. 

\subsection{Classical approach to scattering by the two-atom system}

Let us consider first the atomic system of the previous sections, and let us study its optical response from classical approach. To this end,  we compute their effective electrical polarizabities and apply the  classical formulas derived in the previous subsections.  Following Ref.\cite{myPRA}, the polarizabilities of the active atom, $\alpha_{A}$, and passive atom, $\alpha_{B}$ derive from their response to an external monochromatic field, i.e., $\langle \mathbf{d}_{A,B}(\omega)\rangle=\alpha_{A,B}(\omega)\cdot\mathbf{E}_{0}(\omega)$,
\begin{equation}
	\alpha_{A}= \frac{\mathcal{P}-\gamma}{\Gamma} \frac{\boldsymbol{\mu} \boldsymbol{\mu}}{\hbar(\omega-\omega_{0}+i \Gamma/2))},\quad\alpha_{B}= \frac{\boldsymbol{\mu} \boldsymbol{\mu}}{\hbar(\omega_{0}-\omega-i \gamma/2))}.
\end{equation}

\begin{figure}
	\begin{center}
		\includegraphics[width=135mm,angle=0,clip]{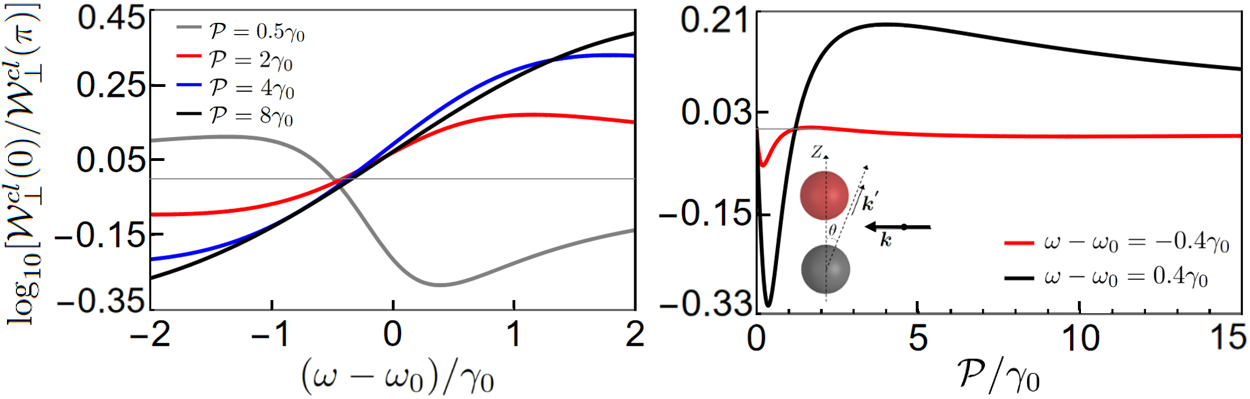}
		\caption{Graphical representation of the log of the ratio of the scattered power towards the active atom and towards the passive one for frontal incidence, according to the classical approach [Eq.(\ref{wclperp})], $\log_{10}{[\mathcal{W}^{cl}_{\perp}(0)/\mathcal{W}^{cl}_{\perp}(\pi)]}$, for fixed interatomic distance $kR=2$, and non-radiative linewidth $\gamma_{nr}=0.2\gamma_{0}$. Left panel: In terms of the detuning $\omega-\omega_{0}$, with several pump rates, $\mathcal{P}=(0.5,2,4,8)\gamma_{0}$. Right panel:  In terms of the pump rate $\mathcal{P}$, with detuning  $\omega-\omega_{0}=0.4\gamma_{0}$ (black line) and $\omega-\omega_{0}=-0.4\gamma_{0}$ (red line).} \label{fig21}
	\end{center}
\end{figure}

In Fig.\ref{fig21} we represent graphically the ratio between the logs of the scattered power towards the active atom ($\theta=0$) and towards the passive one ($\theta=\pi$) for frontal illumination, for interatomic distance $kR=2$, in terms of the detuning with respect to the resonant frequency (left panel) and the pump rate (right panel). In contrast to the results of the quantum approach --cf. Fig.\ref{fig3}, the level of asymmetry is smaller in the classical case, and its direction with respect to the sign of detuning seems the opposite. Also, the symmetry is not restored close to the resonance but for negative detuning. 

\begin{figure}
	\begin{center}
		\includegraphics[width=135mm,angle=0,clip]{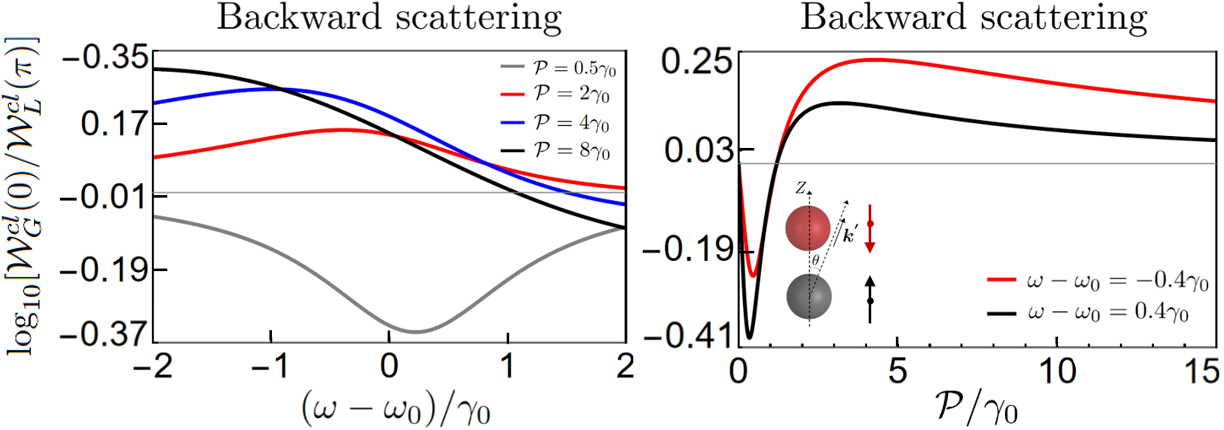}
		\caption{Left panel: graphical representation of the logarithm of the ratio between the backward-scattered powers for side illumination from the gain and loss sides, according to the classical approach [Eq.(\ref{Wclside})], $\log_{10}{[\mathcal{W}_{G}^{cl}(0)/\mathcal{W}_{L}^{cl}(\pi)]}$, in terms of the detuning $\omega-\omega_{0}$.  The interatomic distance is taken at $kR=2$, the non-radiative linewidth is $\gamma_{nr}=0.2\gamma_{0}$, and several pump rates are considered, $\mathcal{P}=(0.5,2,4,8)\gamma_{0}$. Right panel: \emph{Idem} in terms of the detuning $\mathcal{P}$, for fixed detuning, $\omega-\omega_{0}=-0.4\gamma_{0}$ (red line), $\omega-\omega_{0}=0.4\gamma_{0}$ (black line).} \label{fig22}
	\end{center}
\end{figure}

In Fig.\ref{fig22},  we represent the log of the ratio between the backward-scattered powers for side lighting from the gain and loss sides, $\log_{10}{[\mathcal{W}_{G}^{cl}(0)/\mathcal{W}_{L}^{cl}(\pi)]}$, for different pump rates (left panel) and opposite detuning at fixed pump (right panel). In contrast to the quantum results --cf. lower panels of Fig.\ref{fig8}-- the asymmetry changes direction at $\mathcal{P}\approx1.2\gamma_{0}$,  nearly irrespective of the sign of the detuning.

\subsection{Scattering by a dimer of metallic nanoparticles}

Let us consider next a classical system made of a dimer of metallic nanoparticles, one of which is pumped. This time, in order to formulate their electrical polarizabilities, we adopt the Drude model of Ref.\cite{ManjavacasPT1} for the relative dielectric constants of the nanoparticles. That is, we take
\begin{equation}
	\epsilon_{L}=\epsilon_{\infty}-\frac{\omega_{p}^{2}}{\omega^{2}+i\omega\gamma_{p}},\quad \epsilon_{G}=\epsilon_{L}-F\frac{\gamma_{0}}{\omega-\omega_{0}+i\gamma_{0}}
\end{equation}
for the passive or loss-side atom (L), and for the active or gain-side atom (G), respectively. In these expressions $\epsilon_{\infty}$, $\hbar\omega_{p}=2$ eV, $\gamma_{p}=0.05\omega_{p}$ for the passive component, and $\hbar\omega_{0}=1$ eV, $\gamma_{0}=0.01\omega_{0}$ for the Lorentzian gain of the active component, which is intended to describe optical pump, with $F$ being a dimensionless parameter proportional to the population inversion of the gain side. Analogously to the atoms, we consider the nanoparticles as electric dipoles with bare polarizabilities
\begin{equation}
	\alpha_{G,L}=4\pi\epsilon_{0}a^{3}\frac{\epsilon_{G,L}-1}{\epsilon_{G,L}+2}.
\end{equation}
In addition, in order to account consistently for the radiative linewidth of the particles we include the radiation-reaction field in the 'renormalized' polarizabilities of each atom \cite{myPRA1EMvacuum},
\begin{equation}
	\alpha_{A} = \frac{\alpha_{G}}{\mathbb{I}+\alpha_{G}\frac{k^{2}}{\epsilon_{0}}i\textrm{Im}{\mathbb{G}}_{AA}},\quad 	\alpha_{B} = \frac{\alpha_{L}}{\mathbb{I}+\alpha_{L}\frac{k^{2}}{\epsilon_{0}}i\textrm{Im}{\mathbb{G}}_{BB}},
\end{equation}
where $\mathbb{G}_{AA}$ and $\mathbb{G}_{BB}$ stand for $\mathbb{G}(\omega;\mathbf{R}_{A},\mathbf{R}_{A})$ and  $\mathbb{G}(\omega;\mathbf{R}_{B},\mathbf{R}_{B})$, respectively.

\begin{figure}
	\begin{center}
		\includegraphics[width=80mm,angle=0,clip]{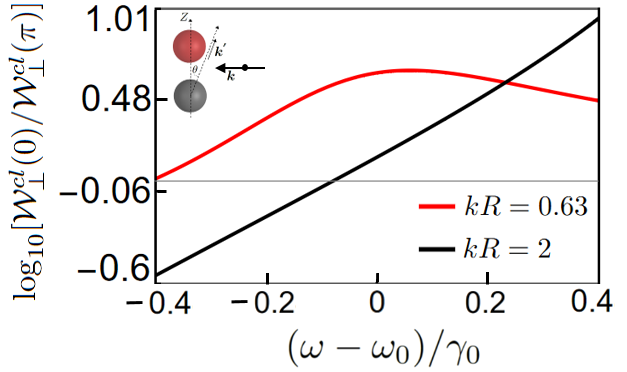}
		\caption{Graphical representation of the ratio between the logs of the scattered power towards the active atom and the passive one for frontal incidence, $\log_{10}{[\mathcal{W}^{cl}_{\perp}(0)/\mathcal{W}^{cl}_{\perp}(\pi)]}$, in terms of the detuning $\omega-\omega_{0}$. The interatomic distance is taken at $kR=0.63,2$, and the pump rate is fixed at $F=0.55$.} \label{fig11}
	\end{center}
\end{figure}

In Fig.\ref{fig11}  we represent graphically the ratio between the logs of the scattered power towards the active atom ($\theta=0$) and towards the passive one ($\theta=\pi$) for frontal illumination, for interatomic distances $kR=0.63$ and $kR=2$, $F\approx0.6$, in terms of the detuning with respect to the resonant frequency. We observe that the scattered power is greater towards the active atom at short distances, regardless of the sign of the detuning \cite{ManjavacasPT1}. However, for $kR=2$, the scattered power is greater towards the active atom for positive values of the detuning, and the tendency gets reversed for negative detuning, contrary to the results found in the quantum system.

\begin{figure}
	\begin{center}
		\includegraphics[width=80mm,angle=0,clip]{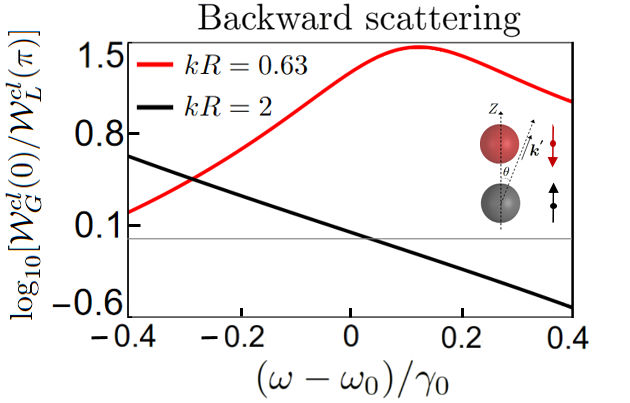}
		\caption{Graphical representation of the logarithm of the ratio between the backward-scattered powers for side illumination from the gain and loss sides,  $\log_{10}{[\mathcal{W}^{cl}_{G}(0)/\mathcal{W}^{cl}_{L}(\pi)]}$, in terms of the detuning $\omega-\omega_{0}$. The interatomic distance is taken at $kR=0.63,2$, and the pump rate is fixed at $F=0.55$.} \label{fig12}
	\end{center}
\end{figure}
In Fig.\ref{fig12}  we represent graphically the ratio between the back-scattered power for G-lighting and for L-lighting, for the interatomic distances $kR=0.63$ and $kR=2$, $F\approx0.6$, in terms of the detuning with respect to the resonant frequency. We observe that backscattering is greater for G-lighting at short distances, regardless of the sign of the detuning \cite{ManjavacasPT1}. However, same as for frontal illumination for $kR=2$, the back-scattered power is greater for L-lighting for positive values of the detuning, and the tendency gets reversed for negative detuning, which is the opposite to the results found in the quantum system.

\section{Discussion and Conclusions}\label{lasec7}

We have performed a quantum analysis of the directionality of the scattered radiation by a binary atomic system made of identical atoms, one of which is excited by an incoherent pump. Our results show that angle-resolved scattering constitutes a powerful probe of the microscopic processes underlying the optical response of interacting quantum emitters.  By resolving the emitted field in space, the microscopic mechanisms associated with photon exchange, interference, and reciprocity become directly identifiable. Thus, on the one hand, we have interpreted the quantum processes in terms of the classical ones found in analogous classical systems. On the other hand, our analysis has revealed distinctive features  of the quantum approach compared with the results obtained within a semiclassical approach, most notably those related to reciprocity. 

To these ends, we have identified all the quantum processes which contribute to scattering at leading order. We have distinguished two kinds of scattering processes, namely, each-atom scattering which results from the interference of pairs of photons emitted by the same atom, and both-atoms scattering, which results from the interference of photons emitted by different atoms. For frontal illumination we find an asymmetry which reaches its maximum value for pump rates of the order of $\gamma_{0}$.  However, the direction of the  imbalance is determined by the sign of the detuning of the probe field with respect to the resonant frequency, being towards the gain side for negative detuning and towards the loss side for positive  detuning for interatomic separations  $kR\lesssim\pi$. The symmetry is nearly restored at exact resonance --see Fig.\ref{fig3}, upper panel. The underlying physical reason for this imbalance is the parity-asymmetry in all both-atoms scattering processes in which photons emitted by different atoms interfere [diagrams (10-16)]. For side lighting we find asymmetry with respect to the side of incidence, for scattering in both forward and backward directions. For interatomic distances $kR\lesssim\pi$, forward scattering is greater for lighting from the loss side, regardless of the sign of the detuning. Hence, the degree of asymmetry is maximum at exact resonance, for $\mathcal{P}\approx2\gamma_{0}$.  On the contrary, the imbalance in backward scattering depends on the sign of the detuning, being generally greater for lighting from the loss side for negative detuning, and being greater for lighting from the gain side for positive detuning --see  Fig.\ref{fig8}. In this case, the underlying physical reason for the imbalance is not only the parity-asymmetry in both-atoms scattering processes that induces a phase-shift due to the interference of photons emitted from different atoms, but also the phase difference between the probe field photons that strike each atom and leaves its mark on the dipole moments induced in either atom. Besides, when the interatomic distance exceeds $kR>\pi$, the imbalance changes direction.  All in all, we conclude that the directional features of scattering in a pumped atomic system which is non-parity symmetric are not generic, for they depend on the interatomic distance and the detuning of the probe field with respect to the resonant frequency. Further, our fully quantum approach reveals the lack of reciprocity in its optical response, which manifests in the asymmetry of forward scattering with respect of the side of incidence. 

\color{black}Several differences have been found with respect to the results obtained through a classical calculation based on linear response theory. In the first place, the classical calculation upon both a binary atomic system and a dimer of metallic nanoparticles--cf. Sec.\ref{lasec6} and Ref.\cite{ManjavacasPT1}, reveals that while for interatomic/interparticle distances greater than a wavelength, $kR\gtrsim1$, the classical calculation predicts a dependence of the asymmetry in the scattered radiation on the sign of the detuning of the probe field with respect to the resonant frequency, in agreement with the quantum result, the positive imbalance is found in the opposite direction with respect to the quantum result, both for frontal and side lighting. 

\color{black}{More importantly, forward scattering for side lighting was found to be independent of the side of incidence in the classical calculations.  On physical grounds, reciprocity is a consequence of the preservation of time-reversal symmetry in the classical optical response of both systems, and despite the fact that the effective polarizability of the active component in the atomic system can be regarded as non-linear since it incorporates the action of the electric pump field. On the basis of the classical calculation, reciprocity results from the compensation of the phase difference between the probe field photons that strike each atom/nanoparticle  with the phase-shift that originates from the interference  of photons emitted from different atoms/particles.  
In contrast, this compensation is only partial in the case of the quantum calculation, as a result of which forward scattering is also asymmetric in the atomic system. 
On the basis of the quantum calculation, this is due to the presence of two-photon intermediate states in some quantum processes.  More specifically, it results from the contribution of the diagrams which starts with the active atom in its excited state  $\tilde{e}$ and possess  two-photon intermediate states [diagrams (7-9), (14-16) in Fig.\ref{fig2}], in which the excited atom emits a photon before absorbing one from the probe field. On physical grounds, we surmise that non-reciprocity must be the result of the violation of time-reversal symmetry along with parity violation. Since no T-violatintg terms are present in the Hamitlonian of the system, such a violation must result from the stationary state of the system, that is, the steady statistical mixure of populations  $\tilde{g}$ and $\tilde{e}$ in the active atom --this is also at the root of the existence of non-reciprocal forces in excited atomic systems \cite{Non-resMe}. 
Thus, for instance, in all the scattering processes, the probe field induces only the transition $g\rightarrow e$ along the time, irrespective of whether the initial state of the active atom is $\tilde{g}$ or $\tilde{e}$. The latter is the case of the aforementioned diagrams with two-photon intermediate states.\color{black}

Let us finish by pointing out that the theoretical framework presented here will provide a useful basis for the analysis and design of more complex light–matter platforms in which directional scattering and collective optical effects play a central role.

\ack
We gratefully acknowledge helpful discussions with Alejandro Manjavacas.  
L.A. acknowledges financial support from the Spanish Ministry of Science, Innovation and Universities (MICIU) through an FPU fellowship (ref. FPU2022-00817). M.D. acknowledges financial support from the Institut Fran\c{c}ais d'Espagne, program Chercheurs Confirm\'es. The work of L.A. was also supported by the project QCAYLE (NextGenerationEU funds, PRTRC17.I1).\\

\noindent\textbf{Data availability statement}

\noindent All the numerical and graphical results that support the findings of this study have been included in the figures. They have been produced using Wolfram Mathematica.\\\\ 

\appendix

In the following appendices we compile the  mathematical expressions of the power associated to some of the diagrams which appear in the main text, for both frontal illumination and side lighting.

\section{Each-atom scattering: scattering power of diagram (6)$+$h.c.}\label{app1}

As an illustrative example of the computation of the scattering power associated to each-atom collective processes, we take diagram (6) and its h.c. The initial and final states in those diagrams are, respectively,
\begin{eqnarray}
\ket{\Psi_{6}^{0}}&=\sqrt{\frac{\gamma}{\Gamma}}\; \ket{N_{\textbf{k}, \boldsymbol{\epsilon}}; \;\tilde{g}, \;g}\nonumber\\
\ket{\Psi_{6}^{f}}&=\sum_{\textbf{k}^{''}, \boldsymbol{\epsilon}^{''}}\ket{(N-1)_{\textbf{k}, \boldsymbol{\epsilon}};\;1_{\textbf{k}^{''}, \boldsymbol{\epsilon}^{''}}; \;\tilde{g}, \;g},\nonumber%
\end{eqnarray}
and, at any intermediate time $\tau$, $\ket{\Psi_{6}(\tau)}=\mathbb{U}(t) \ket{\Psi_{6}^{0}}$. 

The expression for the total scattering power associated to diagram (6) and its h.c. reads, 
	\begin{eqnarray}
\fl	&W^{(6+h.c.)}_{sc}
	 =\frac{\gamma}{\Gamma} \frac{d}{dt}
	\langle\Psi_{8}(t)|\Psi_{8}^{f}\rangle \langle\Psi_{8}^{f}|H_{EM}|\Psi_{8}^{f}\rangle \langle\Psi_{8}^{f}|\Psi_{8}(t)\rangle \; +  \; h.c. \\
\fl	& = \frac{\gamma}{\Gamma}  \frac{d}{dt}\; \left[ \sum_{\boldsymbol{k}',\boldsymbol{\epsilon}'} \sum_{\boldsymbol{k}^{''},\boldsymbol{\epsilon}^{''}} \int_{0}^{t} d\tau \int_{0}^{\tau} d\tau' \int_{0}^{\tau'} d\tau''  \int_{0}^{\tau''} d\tau'''  \int_{0}^{t} d\tilde{\tau} \int_{0}^{\tilde{\tau}} d \tilde{\tau}' \bra{N_{\mathbf{\textbf{k}, \boldsymbol{\epsilon}}}; \;\tilde{g}, \;g} \mathbb{U}_{0}^{\dagger}(\tau''' )\right.\nonumber\\
\fl &\times\left.\ket{N_{\textbf{k}, \boldsymbol{\epsilon}}; \;\tilde{g}, \;g} \bra{N_{\textbf{k}, \boldsymbol{\epsilon}}; \;\tilde{g}, \;g}\mathbf{d}\cdot\mathbf{E}_{\mathbf{\textbf{k}, \boldsymbol{\epsilon}}}^{(-)}(\mathbf{R}_{A})\ket{(N-1)_{\textbf{k}, \boldsymbol{\epsilon}}; \;\tilde{e}, \;g}
	\bra{(N-1)_{\textbf{k}, \boldsymbol{\epsilon}}; \;\tilde{e}, \;g}\right.\nonumber\\ 
\fl &\times\left.	e^{-\frac{\Gamma}{2}(\tau''-\tau''')} \mathbb{U}_{0}^{\dagger}(\tau''-\tau''') \ket{(N-1)_{\textbf{k}, \boldsymbol{\epsilon}}; \;\tilde{e}, \;g}\bra{(N-1)_{\textbf{k}, \boldsymbol{\epsilon}}; \;\tilde{e}, \;g}
\mathbf{d}\cdot\mathbf{E}_{\mathbf{\textbf{k}', \boldsymbol{\epsilon}'}}^{(+)}(\mathbf{R}_{A})\right.\nonumber\\
\fl &\times\left. \ket{(N-1)_{\textbf{k}, \boldsymbol{\epsilon}}; 1_{\textbf{k}', \boldsymbol{\epsilon}'}; \;g, \;g} \bra{(N-1)_{\textbf{k}, \boldsymbol{\epsilon}}; 1_{\textbf{k}', \boldsymbol{\epsilon}'}; \;g, \;g}\mathbb{U}_{0}^{\dagger}(\tau'-\tau'') \ket{(N-1)_{\textbf{k}, \boldsymbol{\epsilon}}; 1_{\textbf{k}', \boldsymbol{\epsilon}'}; \;g, \;g}\right.\nonumber\\
\fl &\times\left.  \bra{(N-1)_{\textbf{k},\boldsymbol{\epsilon}}; 1_{\textbf{k}', \boldsymbol{\epsilon}'}; \;g, \;g}\mathbf{d}\cdot\mathbf{E}_{\mathbf{\textbf{k}', \boldsymbol{\epsilon}'}}^{(-)}(\mathbf{R}_{B})\ket{(N-1)_{\textbf{k}, \boldsymbol{\epsilon}};  \;g, \;e} \bra{(N-1)_{\textbf{k}, \boldsymbol{\epsilon}}; \;g, \;e} e^{-\frac{\gamma}{2}}(\tau-\tau') \right.\nonumber \\
\fl		& \times\left. \mathbb{U}_{0}^{\dagger}(\tau-\tau')\bra{(N-1)_{\textbf{k},\boldsymbol{\epsilon}}; \;g, \;e}\mathbf{d}\cdot\mathbf{E}_{\mathbf{\textbf{k}^{''}, \boldsymbol{\epsilon}^{''}}}^{(+)}(\mathbf{R}_{B}) \ket{(N-1)_{\textbf{k}, \boldsymbol{\epsilon}}; 1_{\textbf{k}^{''}, \boldsymbol{\epsilon}^{''}}; \;g, \;g} \bra{(N-1)_{\textbf{k}, \boldsymbol{\epsilon}}; 1_{\textbf{k}^{''}, \boldsymbol{\epsilon}^{''}}; \;g, \;g }\right.\nonumber\\
\fl		& \times\left. \mathbb{U}_{0}^{\dagger}(t-\tau)\; a_{\textbf{k}^{''}, \boldsymbol{\epsilon}^{''}}^{\dagger} a_{\textbf{k}^{''}, \boldsymbol{\epsilon}^{''}} \; \mathbb{U}_{0}(t-\tilde{\tau})\ket{(N-1)_{\textbf{k}, \boldsymbol{\epsilon}}; 1_{\textbf{k}^{''}, \boldsymbol{\epsilon}^{''}}; \;\tilde{g}, \;g}\bra{(N-1)_{\textbf{k}, \boldsymbol{\epsilon}}; 1_{\textbf{k}^{''}, \boldsymbol{\epsilon}^{''}}; \;\tilde{g}, \;g}\right. \nonumber \\
\fl		& \times\left. \mathbf{d}\cdot\mathbf{E}_{\mathbf{\textbf{k}^{''}, \boldsymbol{\epsilon}^{''}}}^{(-)}(\mathbf{R}_{B}) \ket{(N-1)_{\textbf{k}, \boldsymbol{\epsilon}};  1_{\textbf{k}^{''}, \boldsymbol{\epsilon}^{''}}; \;\tilde{g}, \;e} \bra{(N-1)_{\textbf{k}, \boldsymbol{\epsilon}}; 1_{\textbf{k}^{''}, \boldsymbol{\epsilon}^{''}}; \;\tilde{g}, \;e} \mathbb{U}_{0}(\tilde{\tau}-\tilde{\tau}') e^{-\frac{\gamma}{2}(\tilde{\tau}-\tilde{\tau}')}\right.\nonumber\\
\fl		& \times \left. \ket{(N-1)_{\textbf{k}, \boldsymbol{\epsilon}};  \;\tilde{g}, \;e} \left. \bra{(N-1)_{\textbf{k}, \boldsymbol{\epsilon}}; \;\tilde{g}, \;e} \mathbf{d}\cdot\mathbf{E}_{\mathbf{\textbf{k}, \boldsymbol{\epsilon}}}^{(+)}(\mathbf{R}_{B})\ket{N_{\textbf{k}, \boldsymbol{\epsilon}}; \;\tilde{g}, \;g}
	\bra{N_{\textbf{k}, \boldsymbol{\epsilon}}; \;\tilde{g}, \;g} \mathbb{U}_{0}(\tilde{\tau}')\ket{N_{\textbf{k}, \boldsymbol{\epsilon}}; \;\tilde{g}, \;g} \right]\right.\nonumber\\
	\fl & +\left. c.c. \;, \;\;\; \gamma t \gg 1.\right.\nonumber
	\end{eqnarray}

	\subsection{Frontal illumination}
	
	For frontal illumination the probe field strikes each atom with the same phase, and the product of the amplitudes associated to the interaction factors of the probe-field with each atom results in a prefactor $\hbar^{2}\Omega_{0}^{2}/4$. In the above equation the photons of momentum $\mathbf{k}'$ and polarization vector $\boldsymbol{\epsilon}'$ are those which fly between both atoms, mediate their mutual interaction and induce a dipole moment on atom $B$. They  must be integrated out. In contrast, the photons of momentum $\mathbf{k}''$ and polarization vector $\boldsymbol{\epsilon}^{''}$ are the ones emitted from atom $B$. In the calculation of the differential scattering power we must sum over all their polarization states, integrate over all their frequencies and leave free the integration over the polar angle. Lastly, performing the time integrals and the integral upon the orientations of the photons that mediate the interatomic interaction, we arrive at 	
	\begin{eqnarray}
\fl	&W^{(6+h.c.)}_{\perp}
	 = \frac{ \Omega_{0}^{2} \omega}{16 \hbar\epsilon_{0}^{2} c^{5}} \frac{\gamma}{\Gamma} \int_{0}^{4\pi} \frac{d\Theta_{k^{''}}}{4\pi} \boldsymbol{\mu}\cdot  (\delta_{ij}-\mathbf{\hat{k}^{''}}_{i}\mathbf{\hat{k}^{''}}_{j}) \cdot\boldsymbol{\mu}\nonumber\\
\fl &\times	\frac{d}{dt} \left[\int_{0}^{\infty} \frac{d\omega^{'}}{\pi} \omega^{'2} \boldsymbol{\mu}\cdot \textrm{Im}\mathbb{G}(\omega^{'};R) \cdot\boldsymbol{\mu}  \int_{0}^{\infty} \frac{d\omega^{''}}{\pi} \omega^{''3} e^{-i\omega^{''}r \cos{\theta} }\right. \nonumber \\
\fl	& \times \left. \frac{(e^{-i\omega t} - e^{-i\omega^{''} t})[(\omega^{'}-\omega^{''})(\omega^{'}-\omega_{0}-i\frac{\gamma}{2})(e^{i\omega t} - e^{i\omega^{''} t})  -  (\omega-\omega^{''})(\omega-\omega_{0}-i\frac{\gamma}{2})(e^{i\omega^{'} t} - e^{i\omega^{''} t})]}{(\omega-\omega_{0}+i\frac{\gamma}{2})(\omega^{''}-\omega)(\omega-\omega_{0}-i\frac{\Gamma}{2})(\omega-\omega_{0}-i\frac{\gamma}{2})(\omega^{'}-\omega_{0}-i\frac{\gamma}{2})(\omega-\omega^{'})(\omega^{'}-\omega^{''})(\omega^{'}-\omega^{''})} \right]\nonumber\\
\fl	& + c.c.,\quad r \to 0^{+}.
	\end{eqnarray}
	where the integrals over frequencies $\omega'$ and $\omega''$ as well as the integral over solid angles for the emitted photons, $\Theta_{k^{''}}$, have been indicated. Finally, integrating in frequencies and over the azimuth angle of the emitted photons, we are left with a function of the polar angle $\theta$ that we identify with the differential scattering power, 
	\begin{eqnarray}
	\mathcal{W}^{(6+h.c.)}_{\perp}(\theta)
	& = \frac{ \Omega_{0}^{2} \omega^{4} |\boldsymbol{\mu}|^{2}}{16\pi\epsilon_{0} c^{3} } \;(1+\cos^{2}{\theta}) \; \frac{\mathcal{\gamma}}{\Gamma} \;
	\frac{ \left[(\omega-\omega_{0}) \tilde{\Omega}(R)-\frac{\Gamma}{2} \tilde{\Gamma}(R) \right] }{\left[(\omega-\omega_{0})^{2}+\frac{\gamma^{2}}{4}\right] \left[(\omega-\omega_{0})^{2}+\frac{\Gamma^{2}}{4} \right]}.\nonumber%
	\end{eqnarray}
	
	\subsection{Side lighting from the gain side}
	
	For side lighting from the active atom located at $\mathbf{R}_{A}$, the propagation towards the passive atom at $\mathbf{R}_{B}=\mathbf{R}_{A}-\mathbf{R}$ generates a phase difference between the amplitudes of the probe field that strikes each atom. The product of those amplitudes in diagram (6) results in  $e^{i\mathbf{k}\cdot(\mathbf{R}_{A}-\mathbf{R})}e^{-i\mathbf{k}\cdot\mathbf{R}_{A}}=e^{-i\mathbf{k}\cdot\mathbf{R}}=e^{ikR}$, with $\mathbf{R}$ and $\mathbf{k}$ being antiparallel to each other. Hence, the product of the amplitudes associated to the interaction factors of the probe-field with each atom results in a prefactor $\hbar^{2}\Omega_{0}^{2}e^{ikR}/4$. Again,   performing the time integrals and the integral upon the orientations of the photons that mediate the interatomic interaction in the equation for $W^{(6+h.c.)}_{sc}$, we arrive at	
	\begin{eqnarray}
\fl	&W^{(6+h.c.)}_{G}
	 = \frac{ \Omega_{0}^{2} \omega e^{ikR}}{16 \hbar\epsilon_{0}^{2} c^{5}} \frac{\gamma}{\Gamma} \int_{0}^{4\pi} \frac{d\Theta_{k^{''}}}{4\pi} \boldsymbol{\mu}\cdot  (\delta_{ij}-\mathbf{\hat{k}^{''}}_{i}\mathbf{\hat{k}^{''}}_{j}) \cdot\boldsymbol{\mu}\nonumber\\
\fl &\times	\frac{d}{dt} \left[\int_{0}^{\infty} \frac{d\omega^{'}}{\pi} \omega^{'2} \boldsymbol{\mu}\cdot \textrm{Im}\mathbb{G}(\omega^{'};R) \cdot\boldsymbol{\mu}  \int_{0}^{\infty} \frac{d\omega^{''}}{\pi} \omega^{''3} e^{-i\omega^{''}r \cos{\theta} } \right. \nonumber\\
\fl	& \times \left. \frac{(e^{-i\omega t} - e^{-i\omega^{''} t})[(\omega^{'}-\omega^{''})(\omega^{'}-\omega_{0}-i\frac{\gamma}{2})(e^{i\omega t} - e^{i\omega^{''} t})  -  (\omega-\omega^{''})(\omega-\omega_{0}-i\frac{\gamma}{2})(e^{i\omega^{'} t} - e^{i\omega^{''} t})]}{(\omega-\omega_{0}+i\frac{\gamma}{2})(\omega^{''}-\omega)(\omega-\omega_{0}-i\frac{\Gamma}{2})(\omega-\omega_{0}-i\frac{\gamma}{2})(\omega^{'}-\omega_{0}-i\frac{\gamma}{2})(\omega-\omega^{'})(\omega^{'}-\omega^{''})(\omega^{'}-\omega^{''})} \right]\nonumber\\
\fl	& + c.c.,\quad r \to 0^{+}.
	\end{eqnarray}
	Integrating in frequencies and over the azimuth angle of the emitted photons, we are left with
	\begin{eqnarray}
\fl	\mathcal{W}^{(6+h.c.)}_{G}(\theta)
	=& \frac{ \Omega_{0}^{2} \omega^{4} |\boldsymbol{\mu}|^{2}}{16 \pi \epsilon_{0} c^{3} } \; (1+\cos^{2}{\theta}) \; \frac{\gamma}{\;\Gamma}\Bigl[ \frac{[(\omega-\omega_{0})\cos{kR}-\frac{\Gamma}{2}\sin{kR}]\tilde{\Omega}(R)}{\left[(\omega-\omega_{0})^{2}+\frac{\gamma^{2}}{4}\right] \left[(\omega-\omega_{0})^{2}+\frac{\Gamma^{2}}{4}\right]}\nonumber\\
\fl	&-\frac{[\frac{\Gamma}{2}\cos{kR}+(\omega-\omega_{0})\sin{kR}]\tilde{\Gamma}(R)}{\left[(\omega-\omega_{0})^{2}+\frac{\gamma^{2}}{4}\right] \left[(\omega-\omega_{0})^{2}+\frac{\Gamma^{2}}{4}\right]}\Bigr].
	\end{eqnarray}
	
	\subsection{Side lighting from the loss side}
	
	For side lighting from the passive atom located at $\mathbf{R}_{B}$, the propagation towards the active atom at $\mathbf{R}_{A}=\mathbf{R}_{B}+\mathbf{R}$ generates a phase difference between the  amplitudes of the probe field that strikes each atom. The product of those amplitudes in diagram (6) results in   $e^{i\mathbf{k}\cdot\mathbf{R}_{B}}e^{-i\mathbf{k}\cdot(\mathbf{R}_{B}+\mathbf{R})}=e^{-i\mathbf{k}\cdot\mathbf{R}}=e^{-ikR}$, with $\mathbf{R}$ and $\mathbf{k}$ being parallel to each other. Hence, the product of the amplitudes associated to the interaction factors of the probe-field with each atom results in a prefactor $\hbar^{2}\Omega_{0}^{2}e^{-ikR}/4$. Again,   performing the time integrals and the integral upon the orientations of the photons that mediate the interatomic interaction in the equation for $W^{(6+h.c.)}_{sc}$, we arrive at
	\begin{eqnarray}
\fl	&W^{(6+h.c.)}_{L}
	 = \frac{ \Omega_{0}^{2} \omega e^{-ikR}}{16 \hbar\epsilon_{0}^{2} c^{5}} \frac{\gamma}{\Gamma} \int_{0}^{4\pi} \frac{d\Theta_{k^{''}}}{4\pi} \boldsymbol{\mu}\cdot  (\delta_{ij}-\mathbf{\hat{k}^{''}}_{i}\mathbf{\hat{k}^{''}}_{j}) \cdot\boldsymbol{\mu} \nonumber\\
\fl	& \times \frac{d}{dt} \left[\int_{0}^{\infty} \frac{d\omega^{'}}{\pi} \omega^{'2} \boldsymbol{\mu}\cdot \textrm{Im}\mathbb{G}(\omega^{'};R) \cdot\boldsymbol{\mu}  \int_{0}^{\infty} \frac{d\omega^{''}}{\pi} \omega^{''3} e^{-i\omega^{''}r \cos{\theta} } \right.\nonumber \\
\fl	& \times \left. \frac{(e^{-i\omega t} - e^{-i\omega^{''} t})[(\omega^{'}-\omega^{''})(\omega^{'}-\omega_{0}-i\frac{\gamma}{2})(e^{i\omega t} - e^{i\omega^{''} t})  -  (\omega-\omega^{''})(\omega-\omega_{0}-i\frac{\gamma}{2})(e^{i\omega^{'} t} - e^{i\omega^{''} t})]}{(\omega-\omega_{0}+i\frac{\gamma}{2})(\omega^{''}-\omega)(\omega-\omega_{0}-i\frac{\Gamma}{2})(\omega-\omega_{0}-i\frac{\gamma}{2})(\omega^{'}-\omega_{0}-i\frac{\gamma}{2})(\omega-\omega^{'})(\omega^{'}-\omega^{''})(\omega^{'}-\omega^{''})} \right] \nonumber\\
\fl	& + c.c.,\quad r \to 0^{+}.
	\end{eqnarray}
	
	Integrating in frequencies and over the azimuth angle of the emitted photons, we are left with
	\begin{eqnarray}
\fl	\mathcal{W}^{(6+h.c.)}_{L}(\theta)
	&= \frac{ \Omega_{0}^{2} \omega^{4} |\boldsymbol{\mu}|^{2}}{16 \pi \epsilon_{0} c^{3} } \; (1+\cos^{2}{\theta}) \; \frac{\gamma}{\;\Gamma} \;\left[ \frac{[(\omega-\omega_{0})\cos{kR}+\frac{\Gamma}{2}\sin{kR}]\tilde{\Omega}(R)}{\left[(\omega-\omega_{0})^{2}+\frac{\gamma^{2}}{4}\right] \left[(\omega-\omega_{0})^{2}+\frac{\Gamma^{2}}{4}\right]}\right. \nonumber\\
\fl	&-\left. \frac{[\frac{\Gamma}{2}\cos{kR}-(\omega-\omega_{0})\sin{kR}]\tilde{\Gamma}(R)}{\left[(\omega-\omega_{0})^{2}+\frac{\gamma^{2}}{4}\right] \left[(\omega-\omega_{0})^{2}+\frac{\Gamma^{2}}{4}\right]}\right].
	\end{eqnarray}

\section{Both-atoms scattering: scattering power of diagrams (11) $+$h.c.}\label{app2}

In the first place, as an illustrative example of the computation of the scattering power associated to both-atoms  processes in which the dipole moments on both atoms are directly induced by the probe field, we take diagram (11) and its h.c. The initial and final states in those diagrams are, respectively,
\begin{eqnarray}
\ket{\Psi_{11}^{0}}&=\sqrt{\frac{\mathcal{P}}{\Gamma}}\; \ket{N_{\textbf{k}, \boldsymbol{\epsilon}}; \;\tilde{e}, \;g}\nonumber\\
\ket{\Psi_{11}^{f}}&=\sum_{\textbf{k}', \boldsymbol{\epsilon}'}\ket{(N-1)_{\textbf{k}, \boldsymbol{\epsilon}};\;1_{\textbf{k}', \boldsymbol{\epsilon}'}; \;\tilde{e}, \;g},\nonumber%
\end{eqnarray}
and, at any intermediate time $\tau$, $\ket{\Psi_{11}(\tau)}=\mathbb{U}(t) \ket{\Psi_{11}^{0}}$.

The expression for the total scattering power associated to diagram (11) and its h.c. reads, 
	\begin{eqnarray}
\fl	&\mathcal{W}^{(11+h.c.)}_{sc}
	 =\frac{\mathcal{P}}{\Gamma} \frac{d}{dt}
	\bra{\Psi_{11}(t)}{\Psi_{11}^{f}}\rangle \bra{\Psi_{11}^{f}}H_{EM}\ket{\Psi_{11}^{f}} \langle\Psi_{11}^{f} \ket{\Psi_{11}(t)}\; + \;h.c.\nonumber\\
\fl	& = \frac{\mathcal{P}}{\Gamma} \frac{d}{dt}\left[\sum_{\boldsymbol{k}',\boldsymbol{\epsilon}'} \int_{0}^{t} d\tilde{\tau}  \int_{0}^{\tilde{\tau}} d \tilde{\tau}'  \int_{0}^{t} d\tau \int_{0}^{\tau} d \tau' \bra{N_{\mathbf{\textbf{k}, \boldsymbol{\epsilon}}}; \;\tilde{e}, \;g} \mathbb{U}_{0}^{\dagger}(\tilde{\tau}' )\ket{N_{\textbf{k}, \boldsymbol{\epsilon}}; \;\tilde{e}, \;g} \bra{N_{\textbf{k}, \boldsymbol{\epsilon}}; \;\tilde{e}, \;g} \right.\nonumber\\
\fl	& \times \; \mathbf{d}\cdot\mathbf{E}_{\mathbf{\textbf{k}, \boldsymbol{\epsilon}}}^{(-)}(\mathbf{R}_{B})\ket{(N-1)_{\textbf{k}, \boldsymbol{\epsilon}}; \;\tilde{e}, \;e}
	\bra{(N-1)_{\textbf{k}, \boldsymbol{\epsilon}}; \;\tilde{e}, \;e} e^{-\frac{\gamma}{2}(\tilde{\tau}-\tilde{\tau}')} \mathbb{U}_{0}^{\dagger}(\tilde{\tau}-\tilde{\tau}') \ket{(N-1)_{\textbf{k}, \boldsymbol{\epsilon}}; \;\tilde{e}, \;e}\nonumber \\
\fl	& \times \bra{(N-1)_{\textbf{k}, \boldsymbol{\epsilon}}; \;\tilde{e}, \;e}\mathbf{d}\cdot\mathbf{E}_{\mathbf{\textbf{k}, \boldsymbol{\epsilon}}}^{(+)}(\mathbf{R}_{B}) \ket{(N-1)_{\textbf{k}', \boldsymbol{\epsilon}'}; 1_{\textbf{k}', \boldsymbol{\epsilon}'}; \;\tilde{e}, \;g} \bra{(N-1)_{\textbf{k}, \boldsymbol{\epsilon}}; 1_{\textbf{k}', \boldsymbol{\epsilon}'}; \;\tilde{e}, \;g} \nonumber \\
\fl	& \times \mathbb{U}_{0}^{\dagger}(t-\tilde{\tau})\; a_{\textbf{k}', \boldsymbol{\epsilon}'}^{\dagger} a_{\textbf{k}', \boldsymbol{\epsilon}'} \; \mathbb{U}_{0}(t-\tau)\ket{(N-1)_{\textbf{k}, \boldsymbol{\epsilon}}; 1_{\textbf{k}', \boldsymbol{\epsilon}'}; \;\tilde{e}, \;g}\bra{(N-1)_{\textbf{k}, \boldsymbol{\epsilon}}; 1_{\textbf{k}', \boldsymbol{\epsilon}'}; \;\tilde{e}, \;g} \nonumber \\
\fl	& \times \mathbf{d}\cdot\mathbf{E}_{\mathbf{\textbf{k}, \boldsymbol{\epsilon}}}^{(+)}(\mathbf{R}_{A})\ket{N_{\textbf{k}, \boldsymbol{\epsilon}};  1_{\textbf{k}', \boldsymbol{\epsilon}'}; \;g, \;g} \bra{N_{\textbf{k}, \boldsymbol{\epsilon}}; 1_{\textbf{k}', \boldsymbol{\epsilon}'}; \;g, \;g} \mathbb{U}_{0}(\tau-\tau') e^{-\frac{\Gamma}{2}(\tau-\tau')} \ket{N_{\textbf{k}, \boldsymbol{\epsilon}}; 1_{\textbf{k}', \boldsymbol{\epsilon}'};  \;g, \;g}\nonumber \\
\fl	& \times\left. \bra{N_{\textbf{k}, \boldsymbol{\epsilon}}; 1_{\textbf{k}', \boldsymbol{\epsilon}'}; \;g, \;g} \mathbf{d}\cdot\mathbf{E}_{\mathbf{\textbf{k}', \boldsymbol{\epsilon}'}}^{(-)}(\mathbf{R}_{A})\ket{N_{\textbf{k}, \boldsymbol{\epsilon}}; \;\tilde{e}, \;g}
	\bra{N_{\textbf{k}, \boldsymbol{\epsilon}}; \;\tilde{e}, \;g} \mathbb{U}_{0}(\tau')\ket{N_{\textbf{k}, \boldsymbol{\epsilon}}; \;\tilde{e}, \;g} \right] + c.c.
	\end{eqnarray}

	\subsection{Frontal illumination}
	
	For frontal illumination the probe field strikes each atom with the same phase, and the product of the amplitudes associated to the interaction factors of the probe-field with each atom results in a prefactor $\hbar^{2}\Omega_{0}^{2}/4$. In the above equation, the photons of momentum $\mathbf{k}'$ and polarization vector $\boldsymbol{\epsilon}'$ that seem to fly between both atoms are  the emitted photons. Actually, they do not fly between the atoms, but are emitted from the passive atom $B$ above the observation time, and from the active atom $A$ below the observation time. Hence, process (11) is the result of the interference between the photons scattered from each atom. In the calculation of the differential scattering power we must sum over all their polarization states, integrate over all their frequencies and leave free the integration over the polar angle. Lastly, summing over all the polarization states and performing the time integrals, we arrive at 
	
	\begin{eqnarray}
\fl	&W^{(11+h.c.)}_{\perp}
	 = \frac{ \Omega_{0}^{2} \omega}{16 \pi\epsilon_{0} c^{3} } \frac{\mathcal{P}}{\Gamma} \int_{0}^{4\pi} \frac{d\Theta_{k^{'}}}{4\pi} \boldsymbol{\mu}\cdot  (\delta_{ij}-\mathbf{\hat{k}^{'}}_{i}\mathbf{\hat{k}^{'}}_{j}) \cdot\boldsymbol{\mu} \\
\fl	& \times \frac{d}{dt} \left[ \int_{0}^{\infty} \frac{d\omega^{'}}{\pi} \omega^{'3} e^{-i \omega^{'}R\cos{\theta}} \times \frac{e^{i\omega t} e^{-i\omega^{'} t} + e^{-i\omega t} e^{i\omega^{'} t}}{(\omega^{'}-\omega_{0}-i\frac{\Gamma}{2})(\omega-\omega_{0}-i\frac{\gamma}{2})(\omega^{'}-\omega)^{2}} \right] + c.c. \;, \;\;\;\; \gamma t \gg 1.\nonumber
	\end{eqnarray}
	
	Integrating in frequencies and over the azimuth angle of the emitted photons, we are left with a function of the polar angle $\theta$ that we identify with the differential scattering power,
	
	\begin{eqnarray}
\fl	\mathcal{W}^{(11+h.c.)}_{\perp}
	&= -\frac{ \Omega_{0}^{2} \omega^{4} |\boldsymbol{\mu}|^{2}}{16\pi\epsilon_{0} c^{3} } \;(1+\cos^{2}{\theta}) \frac{\mathcal{P}}{\Gamma}\left[ 
	\frac{\left[(\omega-\omega_{0})^{2}-\frac{\gamma \Gamma}{4}\right] \cos{(kR \cos{\theta})}}{\left[(\omega-\omega_{0})^{2}+\frac{\gamma^{2}}{4}\right] \left[(\omega-\omega_{0})^{2}+\frac{\Gamma^{2}}{4}\right]}\right.\nonumber\\
\fl &+\left.\frac{\left(\frac{\gamma+\Gamma}{2}\right) (\omega-\omega_{0}) \sin{(kR \cos{\theta})}}{\left[(\omega-\omega_{0})^{2}+\frac{\gamma^{2}}{4}\right] \left[(\omega-\omega_{0})^{2}+\frac{\Gamma^{2}}{4}\right]}\right].\nonumber%
	\end{eqnarray}
	
	\subsection{Side lighting from the gain side}
	
	For side lighting from the active atom located at $\mathbf{R}_{A}$, again the propagation towards the passive atom at $\mathbf{R}_{B}=\mathbf{R}_{A}-\mathbf{R}$ generates a phase difference between the amplitudes of the probe field that strikes each atom. The product of those amplitudes in diagram (11) results in  $e^{i\mathbf{k}\cdot(\mathbf{R}_{A}-\mathbf{R})}e^{-i\mathbf{k}\cdot\mathbf{R}_{A}}=e^{-i\mathbf{k}\cdot\mathbf{R}}=e^{ikR}$, with $\mathbf{R}$ and $\mathbf{k}$ being antiparallel to each other. Hence, the product of the amplitudes associated to the interaction factors of the probe-field with each atom results in a prefactor $\hbar^{2}\Omega_{0}^{2}e^{ikR}/4$. Again,  summing over all the polarization states and performing the time integrals in the equation for $W^{(11+h.c.)}_{sc}$, we arrive at
	
	\begin{eqnarray}
\fl	&W^{(11+h.c.)}_{G}
	 = \frac{ \Omega_{0}^{2} \omega e^{-ikR}}{16 \pi\epsilon_{0} c^{3} } \frac{\mathcal{P}}{\Gamma} \int_{0}^{4\pi} \frac{d\Theta_{k^{'}}}{4\pi} \boldsymbol{\mu}\cdot  (\delta_{ij}-\mathbf{\hat{k}^{'}}_{i}\mathbf{\hat{k}^{'}}_{j}) \cdot\boldsymbol{\mu} \\
\fl	& \times \frac{d}{dt} \left[ \int_{0}^{\infty} \frac{d\omega^{'}}{\pi} \omega^{'3} e^{-i \omega^{'}R\cos{\theta}} \times \frac{e^{i\omega t} e^{-i\omega^{'} t} + e^{-i\omega t} e^{i\omega^{'} t}}{(\omega^{'}-\omega_{0}-i\frac{\Gamma}{2})(\omega-\omega_{0}-i\frac{\gamma}{2})(\omega^{'}-\omega)^{2}} \right] + c.c. \;, \;\;\;\; \gamma t \gg 1.\nonumber
	\end{eqnarray}
	
	Integrating in frequencies and over the azimuth angle of the emitted photons, we are left with 
	\begin{eqnarray}
\fl	\mathcal{W}^{(11+h.c.)}_{G}
	&= -\frac{ \Omega_{0}^{2} \omega^{4} |\boldsymbol{\mu}|^{2}}{16\pi\epsilon_{0} c^{3} } \;(1+\cos^{2}{\theta}) \; \frac{\mathcal{P}}{\Gamma}\left[ \frac{\left[(\omega-\omega_{0})^{2}-\frac{\gamma \Gamma}{4}\right] \cos{(kR+kR \cos{\theta})} }{\left[(\omega-\omega_{0})^{2}+\frac{\gamma^{2}}{4}\right] \left[(\omega-\omega_{0})^{2}+\frac{\Gamma^{2}}{4}\right]}\right.\nonumber\\
\fl	&+\left.\frac{\left(\frac{\gamma+\Gamma}{2}\right) (\omega-\omega_{0}) \sin{(kR+kR \cos{\theta})}}{\left[(\omega-\omega_{0})^{2}+\frac{\gamma^{2}}{4}\right] \left[(\omega-\omega_{0})^{2}+\frac{\Gamma^{2}}{4}\right]}\right].
	\end{eqnarray}
	
	\subsection{Side lighting from the loss side}
	
	For side lighting from the passive atom located at $\mathbf{R}_{B}$, the propagation towards the active atom at $\mathbf{R}_{A}=\mathbf{R}_{B}+\mathbf{R}$ generates a phase difference between the  amplitudes of the probe field that strikes each atom. The product of those amplitudes in diagram (11) results in   $e^{i\mathbf{k}\cdot\mathbf{R}_{B}}e^{-i\mathbf{k}\cdot(\mathbf{R}_{B}+\mathbf{R})}=e^{-i\mathbf{k}\cdot\mathbf{R}}=e^{-ikR}$, with $\mathbf{R}$ and $\mathbf{k}$ being parallel to each other. Hence, the product of the amplitudes associated to the interaction factors of the probe-field with each atom results in a prefactor $\hbar^{2}\Omega_{0}^{2}e^{-ikR}/4$. Summing over all the polarization states and performing the time integrals in the equation for $W^{(11+h.c.)}_{sc}$, we arrive at
	
	\begin{eqnarray}
\fl	&W^{(11+h.c.)}_{L}
	 = \frac{ \Omega_{0}^{2} \omega e^{ikR}}{16 \pi\epsilon_{0} c^{3} } \frac{\mathcal{P}}{\Gamma} \int_{0}^{4\pi} \frac{d\Theta_{k^{'}}}{4\pi} \boldsymbol{\mu}\cdot  (\delta_{ij}-\mathbf{\hat{k}^{'}}_{i}\mathbf{\hat{k}^{'}}_{j}) \cdot\boldsymbol{\mu} \\
\fl	& \times \frac{d}{dt} \left[ \int_{0}^{\infty} \frac{d\omega^{'}}{\pi} \omega^{'3} e^{-i \omega^{'}R\cos{\theta}} \times \frac{e^{i\omega t} e^{-i\omega^{'} t} + e^{-i\omega t} e^{i\omega^{'} t}}{(\omega^{'}-\omega_{0}-i\frac{\Gamma}{2})(\omega-\omega_{0}-i\frac{\gamma}{2})(\omega^{'}-\omega)^{2}} \right] + c.c. \;, \;\;\;\; \gamma t \gg 1.\nonumber
	\end{eqnarray}
	
	Finally, integrating in frequencies and over the azimuth angle of the emitted photons, we are left with 
	\begin{eqnarray}
\fl	\mathcal{W}^{(11+h.c.)}_{L}
	&= -\frac{ \Omega_{0}^{2} \omega^{4} |\boldsymbol{\mu}|^{2}}{16\pi\epsilon_{0} c^{3} } \;(1+\cos^{2}{\theta}) \frac{\mathcal{P}}{\Gamma}\left[
	\frac{\left[(\omega-\omega_{0})^{2}-\frac{\gamma \Gamma}{4}\right] \cos{(-kR+kR \cos{\theta})}}{\left[(\omega-\omega_{0})^{2}+\frac{\gamma^{2}}{4}\right] \left[(\omega-\omega_{0})^{2}+\frac{\Gamma^{2}}{4}\right]}\right.\nonumber\\
\fl &+	\left.\frac{\left(\frac{\gamma+\Gamma}{2}\right) (\omega-\omega_{0}) \sin{(-kR+kR \cos{\theta})}}{\left[(\omega-\omega_{0})^{2}+\frac{\gamma^{2}}{4}\right] \left[(\omega-\omega_{0})^{2}+\frac{\Gamma^{2}}{4}\right]}\right].
	\end{eqnarray}	
	
	\section{Both-atoms scattering: scattering power of diagrams (14) $+$h.c.}\label{app3}
	
	In the first place, as an illustrative example of the computation of the scattering power associated to both-atoms  processes in which only the dipole moment is directly induced by the probe field on one of the atoms, we take diagram (14) and its h.c. The initial and final states in those diagrams are, respectively,
	\begin{eqnarray}
	\ket{\Psi_{14}^{0}}&=\sqrt{\frac{\mathcal{P}}{\Gamma}}\; \ket{N_{\textbf{k}, \boldsymbol{\epsilon}}; \;\tilde{e}, \;g}\nonumber\\
	\ket{\Psi_{14}^{f}}&=\sum_{\textbf{k}'', \boldsymbol{\epsilon}''}\ket{(N-1)_{\textbf{k}, \boldsymbol{\epsilon}};\;1_{\textbf{k}', \boldsymbol{\epsilon}'}; \;\tilde{e}, \;g},\nonumber%
	\end{eqnarray}
	and, at any intermediate time $\tau$, $\ket{\Psi_{14}(\tau)}=\mathbb{U}(t) \ket{\Psi_{14}^{0}}$.
	
	The expression for the total scattering power associated to diagram (14) and its h.c. reads, 
		\begin{eqnarray}
	\fl	&\mathcal{W}^{(14+h.c.)}_{sc}
		 =\frac{\mathcal{P}}{\Gamma} \frac{d}{dt}
		\langle\Psi_{14}(t)|\Psi_{14}^{f}\rangle \langle\Psi_{14}^{f}|H_{EM}|\Psi_{14}^{f}\rangle \langle\Psi_{14}^{f}|\Psi_{14}(t)\rangle \; +  \; h.c.\nonumber\\
	\fl	& = \frac{\mathcal{P}}{\Gamma}  \frac{d}{dt}\; \left[ \sum_{\boldsymbol{k}^{'},\boldsymbol{\epsilon}^{'}} \sum_{\boldsymbol{k}^{''},\boldsymbol{\epsilon}^{''}} \int_{0}^{t} d\tau \int_{0}^{\tau} d\tau^{'} \int_{0}^{\tau^{'}} d\tau^{''}  \int_{0}^{\tau^{''}} d\tau^{'''}  \int_{0}^{t} d\tilde{\tau} \int_{0}^{\tilde{\tau}} d \tilde{\tau}^{'} \bra{N_{\mathbf{\textbf{k}, \boldsymbol{\epsilon}}}; \;\tilde{e}, \;g} \mathbb{U}_{0}^{\dagger}(\tau^{'''} )\ket{N_{\textbf{k}, \boldsymbol{\epsilon}}; \;\tilde{e}, \;g}  \right.\nonumber\\
	\fl	& \times \bra{N_{\textbf{k}, \boldsymbol{\epsilon}}; \;\tilde{e}, \;g}\mathbf{d}\cdot\mathbf{E}_{\mathbf{\textbf{k}^{'}, \boldsymbol{\epsilon}^{'}}}^{(+)}(\mathbf{R}_{A})\ket{N_{\textbf{k}, \boldsymbol{\epsilon}}; 1_{\textbf{k}^{'}, \boldsymbol{\epsilon}^{'}}; \;g, \;g}
		\bra{N_{\textbf{k}, \boldsymbol{\epsilon}}; 1_{\textbf{k}^{'}, \boldsymbol{\epsilon}^{'}}; \;g, \;g} e^{-\frac{\Gamma}{2}(\tau^{''}-\tau^{'''})} \mathbb{U}_{0}^{\dagger}(\tau^{''}-\tau^{'''})\nonumber\\
	\fl	& \times \ket{N_{\textbf{k}, \boldsymbol{\epsilon}};1_{\textbf{k}^{'}, \boldsymbol{\epsilon}^{'}} \;g, \;g}\bra{N_{\textbf{k}, \boldsymbol{\epsilon}};1_{\textbf{k}^{'}, \boldsymbol{\epsilon}^{'}} \;g, \;g}\mathbf{d}\cdot\mathbf{E}_{\mathbf{\textbf{k}^{'}, \boldsymbol{\epsilon}^{'}}}^{(-)}(\mathbf{R}_{B}) \ket{N_{\textbf{k}, \boldsymbol{\epsilon}}; \;g, \;e} \bra{N_{\textbf{k}, \boldsymbol{\epsilon}}; \;g, \;e}e^{-\frac{(\gamma+\Gamma)}{2}(\tau^{'}-\tau^{''})}\nonumber\\ 
	\fl	& \times\mathbb{U}_{0}^{\dagger}(\tau^{'}-\tau^{''}) \ket{N_{\textbf{k}, \boldsymbol{\epsilon}}; \;g, \;e} \bra{N_{\textbf{k},\boldsymbol{\epsilon}}; \;g, \;e}\mathbf{d}\cdot\mathbf{E}_{\mathbf{\textbf{k}^{''}, \boldsymbol{\epsilon}^{''}}}^{(+)}(\mathbf{R}_{B}) \ket{N_{\textbf{k}, \boldsymbol{\epsilon}};1_{\textbf{k}^{''}, \boldsymbol{\epsilon}^{''}}; \;g, \;g} \bra{N_{\textbf{k}, \boldsymbol{\epsilon}};1_{\textbf{k}^{''}, \boldsymbol{\epsilon}^{''}}; \;g, \;g} \nonumber \\
	\fl	& \times e^{-\frac{\Gamma}{2}(\tau-\tau^{'})} \mathbb{U}_{0}^{\dagger}(\tau-\tau^{'}) \ket{N_{\textbf{k}, \boldsymbol{\epsilon}};1_{\textbf{k}^{''}, \boldsymbol{\epsilon}^{''}}; \;g, \;g} \bra{N_{\textbf{k},\boldsymbol{\epsilon}}; 1_{\textbf{k}^{''}, \boldsymbol{\epsilon}^{''}}; \;g, \;g}\mathbf{d}\cdot\mathbf{E}_{\mathbf{\textbf{k}, \boldsymbol{\epsilon}}}^{(-)}(\mathbf{R}_{A})\nonumber\\
	\fl	& \times \ket{(N-1)_{\textbf{k}, \boldsymbol{\epsilon}}; 1_{\textbf{k}^{''}, \boldsymbol{\epsilon}^{''}}; \;e, \;g} \bra{(N-1)_{\textbf{k}, \boldsymbol{\epsilon}}; 1_{\textbf{k}^{''}, \boldsymbol{\epsilon}^{''}}; \;e, \;g }\mathbb{U}_{0}^{\dagger}(t-\tau)  a_{\textbf{k}^{''}, \boldsymbol{\epsilon}^{''}}^{\dagger} a_{\textbf{k}^{''}, \boldsymbol{\epsilon}^{''}} \; \mathbb{U}_{0}(t-\tilde{\tau})\nonumber\\
	\fl	& \times\ket{(N-1)_{\textbf{k}, \boldsymbol{\epsilon}}; 1_{\textbf{k}^{''}, \boldsymbol{\epsilon}^{''}}; \;\tilde{e}, \;g}\bra{(N-1)_{\textbf{k}, \boldsymbol{\epsilon}}; 1_{\textbf{k}^{''}, \boldsymbol{\epsilon}^{''}}; \;\tilde{e}, \;g}\mathbf{d}\cdot\mathbf{E}_{\mathbf{\textbf{k}, \boldsymbol{\epsilon}}}^{(+)}(\mathbf{R}_{A})\ket{N_{\textbf{k}, \boldsymbol{\epsilon}};  1_{\textbf{k}^{''}, \boldsymbol{\epsilon}^{''}}; \;\tilde{g}, \;g}\nonumber\\
	\fl	& \times  \bra{N_{\textbf{k}, \boldsymbol{\epsilon}}; 1_{\textbf{k}^{''}, \boldsymbol{\epsilon}^{''}}; \;\tilde{g}, \;g} \mathbb{U}_{0}(\tilde{\tau}-\tilde{\tau}^{'}) e^{-\frac{\Gamma}{2}(\tilde{\tau}-\tilde{\tau}^{'})} \ket{N_{\textbf{k}, \boldsymbol{\epsilon}};1_{\textbf{k}^{''}, \boldsymbol{\epsilon}^{''}};  \;\tilde{g}, \;g}  \bra{N_{\textbf{k}, \boldsymbol{\epsilon}};1_{\textbf{k}^{''}, \boldsymbol{\epsilon}^{''}}; \;\tilde{g}, \;g}\mathbf{d}\cdot\mathbf{E}_{\mathbf{\textbf{k}^{''}, \boldsymbol{\epsilon}^{''}}}^{(-)}(\mathbf{R}_{A})\nonumber\\
	\fl &\times \left.\ket{N_{\textbf{k}, \boldsymbol{\epsilon}}; \;\tilde{e}, \;g}
	\bra{N_{\textbf{k}, \boldsymbol{\epsilon}}; \;\tilde{e}, \;g} \mathbb{U}_{0}(\tilde{\tau}^{'})\ket{N_{\textbf{k}, \boldsymbol{\epsilon}}; \;\tilde{e}, \;g} \right] + c.c.,\quad \gamma t \gg 1.
		\end{eqnarray}
		Since the probe field strikes only the active atom, the product of the amplitudes associated to the interaction factors of the probe-field with an only atom results in a prefactor $\hbar^{2}\Omega_{0}^{2}/4$ common to both frontal and side illumination. In the above equation, the photons of momentum $\mathbf{k}'$ and polarization vector $\boldsymbol{\epsilon}'$ are those which fly between both atoms, mediate their mutual interaction and induce a dipole moment on atom $B$. They  must be integrated out. In contrast, the photons of momentum $\mathbf{k}''$ and polarization vector $\boldsymbol{\epsilon}^{''}$ are the ones emitted from the passive atom $B$ above the observation time, and from the active atom $A$ below the observation time. Hence, process (14) is the result of the interference between the photons scattered from each atom in diagrams (4) and (9). In the calculation of the differential scattering power we must sum over all their polarization states, integrate over all their frequencies and leave free the integration over the polar angle. Lastly, summing over all the polarization states and performing the time integrals, we arrive at 
		\begin{eqnarray}
	\fl	&\mathcal{W}^{(14+h.c.)}_{sc}
		 = \frac{ \Omega_{0}^{2} \omega}{16 \hbar\epsilon_{0}^{2} c^{5}} \frac{\mathcal{P}}{\Gamma} \int_{0}^{4\pi} \frac{d\Theta_{k^{''}}}{4\pi} \boldsymbol{\mu} \cdot(\delta_{ij}-\mathbf{\hat{k}^{''}}_{i}\mathbf{\hat{k}^{''}}_{j}) \cdot\boldsymbol{\mu}\nonumber\\
	\fl	 &\times \frac{d}{dt} \left[\int_{0}^{\infty} \frac{d\omega^{'}}{\pi} \omega^{'2} \boldsymbol{\mu}\cdot\textrm{Im}\mathbb{G}(\omega^{'};R) \cdot\boldsymbol{\mu}\int_{0}^{\infty} \frac{d\omega^{''}}{\pi} \omega^{''3} e^{-i\omega^{''}R \cos{\theta} }\right.\nonumber\\
	\fl	& \times \left. \frac{(e^{-i\omega t} - e^{-i\omega^{''} t})[(\omega^{'}-\omega^{''})(\omega^{'}-\omega_{0}-i\frac{\gamma}{2})(e^{i\omega t} - e^{i\omega^{''} t})  -  (\omega-\omega^{''})(\omega-\omega_{0}-i\frac{\gamma}{2})(e^{i\omega^{'} t} - e^{i\omega^{''} t})]}{(\omega-\omega_{0}+i\frac{\gamma}{2})(\omega^{''}-\omega)(\omega-\omega_{0}-i\frac{\Gamma}{2})(\omega-\omega_{0}-i\frac{\gamma}{2})(\omega^{'}-\omega_{0}-i\frac{\gamma}{2})(\omega-\omega^{'})(\omega^{'}-\omega^{''})(\omega^{'}-\omega^{''})} \right]\nonumber\\
	\fl & + c.c.
		\end{eqnarray}
		Integrating in frequencies and over the azimuth angle of the emitted photons, we are left with a function of the polar angle $\theta$ that we identify with the differential scattering power and is independent of the side of incidence of the probe field,
		\begin{equation}
\fl		\mathcal{W}^{(14+h.c.)}_{sc}
		 = \frac{ \Omega_{0}^{2} \omega^{4} |\boldsymbol{\mu}|^{2}}{16\pi\epsilon_{0} c^{3} } \;(1+\cos^{2}{\theta}) \; \frac{\mathcal{P}}{\Gamma} \;
		\frac{ 2\left[\sin{(\omega R\cos{\theta})}\tilde{\Omega}(R)  - \cos{(\omega R \cos{\theta})} \tilde{\Gamma}(R) \right] }{(\gamma+\Gamma) \left[(\omega-\omega_{0})^{2}+\frac{\Gamma^{2}}{4} \right]}.\nonumber
		\end{equation}

\end{document}